\documentclass[fleqn,usenatbib]{mnras}

\usepackage[T1]{fontenc}

\DeclareRobustCommand{\VAN}[3]{#2}
\let\VANthebibliography\thebibliography
\def\thebibliography{\DeclareRobustCommand{\VAN}[3]{##3}\VANthebibliography}

\usepackage{graphicx}	
\usepackage{amsmath}	
\usepackage{amssymb}	
\usepackage{caption}
\usepackage{subcaption}

\newcommand\eg{{\it e.g.} }

\newcommand{\Gaia}{\textit{Gaia} }
\newcommand{\Gaias}{\textit{Gaia}} 
\newcommand{\tm}[1]{\textrm{#1}}

\title[Can \textit{Gaia} find planets around white dwarfs? ]{Can \textit{Gaia} find planets around white dwarfs?}

\author[]{
Hannah Sanderson$^{1,2}$\thanks{E-mail: hannah.sanderson@earth.ox.ac.uk}, Amy Bonsor$^{2}$, Alexander Mustill$^{3}$
\\
$^{1}$Department of Earth Sciences, South Parks Road, Oxford, OX1 3AN, UK\\
$^{2}$Institute of Astronomy, University of Cambridge, Madingley Road, Cambridge, CB3 0HA, UK\\
$^{3}$Lund Observatory, Dept. of Astronomy and Theorical Physics, Lund University, Box 43, 22100 Lund, Sweden}

\date{Accepted XXX. Received YYY; in original form ZZZ}

\pubyear{2022}

\begin{document}
\label{firstpage}
\pagerange{\pageref{firstpage}--\pageref{lastpage}}
\maketitle

\begin{abstract} 
The \Gaia spacecraft presents an  unprecedented opportunity to reveal the population of long period (a>1\,au) exoplanets orbiting stars across the H-R diagram, including white dwarfs. White dwarf planetary systems have played an important role in the study of planetary compositions, from their unique ability to provide bulk elemental abundances of planetary material in their atmospheres. Yet, very little is known about the population of planets around white dwarfs.
This paper predicts the population of planets that \Gaia will detect around white dwarfs, evolved from known planets orbiting main-sequence stars. 
We predict that \Gaia will detect $8\pm2$ planets around white dwarfs: $8\pm\,3\%$ will lie inside 3\,au and $40\pm10\,\%$ will be less massive than Jupiter. As surviving planets likely become dynamically detached from their outer systems, those white dwarfs with \Gaia detected planets may not have planetary material in their atmospheres. Comparison between the predicted planet population and that found by \Gaia will reveal the importance of dynamical instabilities and scattering of planets after the main-sequence, as well as whether photoevaporation removes the envelopes of gas giants during their giant branch evolution. 

\end{abstract}

\section{Introduction}
The \Gaia space satellite has provided an unprecedented level of information on about 1.8 billion stars in the Milky Way (and beyond) leading to advances across astrophysics from structure and evolution of the Milky Way to asteroseismology to tests of General Relativity \citep{gaia_collaboration_gaia_2016-1,gaia_collaboration_gaia_2018,klioner_gaia_2021,asteroseismology_gaia_2022}. In providing precision astrometry for all nearby stars, \Gaia will uncover many companions to those stars, including exoplanets. Uniquely this detection method will access stars at all stages of stellar evolution, including white dwarfs. Already, in data release three (DR3), \Gaia astrometric, spectroscopic and radial velocity measurments have brought the number of  known binary systems across the H-R diagram to 800,000 \citep{arenou_gaia_2022}. \Gaia has detected ten exoplanet candidates via radial velocity measurements. Eleven main-sequence exoplanets between 4-20\,$M_J$ have been detected astrometrically: nine have been validated against existing detections and two are new detections. Four white dwarf substellar companion candidates have been detected, including the detection of one super Jupiter candidate around a white dwarf. In future data releases, the reduction in the errors and the longer baseline for measurements will enable detection of many more systems. 

Understanding the planetary systems around white dwarfs is key, because these stars potentially reveal planetary composition. Polluted white dwarfs, white dwarfs with metal lines in their spectra, trace the composition of tidally disrupted planetesimals that have been accreted onto the star \citep{jura_extrasolar_2014}. This can provides an opportunity to study the interiors of rocky exoplanets \citep{harrison_polluted_2018,swan_collisions_2021,zuckerman_aluminumcalcium-rich_2011} and the architecture of planetary systems post-main sequence \citep{mustill_unstable_2018,veras_great_2011}. Although polluted white dwarfs provide signatures of tidally disrupted bodies, little is known about planets orbiting white dwarfs. Despite, extensive transit and direct imaging surveys \citep{fulton_search_2014,hogan_latest_2011,faedi_detection_2011,burleigh_imaging_2002,gould_finding_2008,debes_cool_2005,xu_extreme-ao_2015} a limited number of white dwarf planet candidates have been discovered, including \citet{sigurdsson_young_2003,luhman_discovery_2011,gansicke_accretion_2019,vanderburg_giant_2020,blackman_jovian_2021} which were  discovered prior to DR3. One candidate has a semi-major axis around 3\,au \citep{blackman_jovian_2021}, but the other planet candidates are in extreme environments. Some are very close e.g. orbital period $\sim1$\,day \citep{vanderburg_giant_2020} or close enough to be photoevaporating \citep{gansicke_accretion_2019}. Others are very far ($\sim\,2500$\,au) \citep{luhman_discovery_2011} from their host star or circumbinary around a white dwarf and pulsar \citep{sigurdsson_young_2003}. Detailed observations of planets around white dwarfs are lacking. 

Most known planets orbit host stars that end their lives as white dwarfs, and models suggest that outer planets should survive post-main sequence stellar evolution \citep{veras_post-main-sequence_2016,mustill_foretellings_2012}. The final position of a planet after post-main sequence stellar evolution is determined by the competing effects of mass loss, which moves planets outwards, and tidal forces, which move them inwards. Tidal forces are most significant for massive planets on the AGB, when stars have more extended envelopes \citep[up to several au][]{mustill_foretellings_2012}. Close-in planets may be engulfed, whilst the high luminosity of the AGB star can cause heating and mass loss from the exoplanet atmosphere \citep[photoevaporation][]{veras_post-main-sequence_2016,barker_tidal_2020,bear_evaporation_2011}. In multi-planet systems, these effects are complicated by the influence of planets on each other and resulting instabilities can lead to ejection of planets or star-planet collisions \citep{veras_simulations_2013}.

Although the broad effects are known, detailed modelling of post-main sequence planetary evolution is difficult due to uncertainties in stellar evolution models for the RGB and AGB \citep[e.g.][]{bertolami_evolutionary_2018, matthews_evolved_2018}, different approaches to modelling tidal interactions \citep{ogilvie_tidal_2014}, problems quantifying energy sources that contribute to planet common envelope evolution \citep{ivanova_common_2013}, chaotic evolution of multi-planet systems e.g. \citet{veras_simulations_2013} and the uncertainty in the main-sequence population of wide-orbit planets around WD progenitors. Observationally probing the population of  planets around white dwarfs would provide us with direct evidence of planets that have survived this post-main sequence evolution, which is vital for improving models of this process.

With its high precision astrometry, \Gaia presents an unprecedented opportunity to uncover the population of planets orbiting white dwarfs. Crucially, \Gaia will  bridge the gap between the small (<\,1\,au) semi-major axes of transit surveys and the large (>\,10\,au) semi-major axes of direct imaging surveys. \Gaia is predicted to find  tens of thousands of planets beyond 1\,au \citep{casertano_double-blind_2008,perryman_astrometric_2014,ranalli_astrometry_2018}, including some around white dwarfs. \citet{silvotti_white_2011} show that \Gaia will uncover planets with masses greater than $2M_J$ around the brightest white dwarfs. \Gaia has already demonstrated its potential to find white dwarf exoplanets with the announcement of a new white dwarf planet candidate in DR3: a $\sim9$\,$M_J$ planet with a period of $33.65\pm0.05$ days around WD0141-675 \citep{arenou_gaia_2022}. This is also the first planet candidate detection around a polluted white dwarf. The full catalogue, which will include planets with periods up to $\approx10$ years, will not be available until DR5. Comparison between HIPPARCOS and \Gaia proper motion anomalies can also be used to get obtain a longer baseline \citep[24.75 years between HIPPARCOS and DR3][]{kervella_stellar_2022}. Most white dwarfs were too faint to be observed in detail by HIPPARCOS, but this technique has already hinted at binarity in two white dwarf systems  (LAWD 37, GD 140) \citet{kervella_stellar_2019,kervella_stellar_2022}. 

This paper aims to provide a benchmark with which to assess those planets detected around white dwarfs by \Gaias. The aim is to make predictions for the population of planets that \Gaia will detect around white dwarfs by evolving the population of planets seen around main-sequence stars. Comparison of these predictions with the population of planets that \Gaia detects will probe the importance of additional processes, such as dynamical scattering, survival of common envelope evolution or second generation planet formation. If planets are to arrive in the habitable zone around white dwarfs at around 0.01\,au, a potential avenue for the origin of life, such processes are key \citep{agol_transit_2011,loeb_detecting_2013,kaltenegger_white_2020}. This paper starts by predicting the mass and semi-major axes of planets that \Gaia can detect (\S\ref{methods-probability}, \S\ref{results-probability}). Predictions are then made for the fate of currently detected planets around main-sequence stars (\S\ref{methods-distribution}), which are used to obtain a prediction for the distribution of planets in the \Gaia detection region that have evolved from the main-sequence via tides and stellar mass loss (\S\ref{results-planet-distribution}). By combining the \Gaia detection probabilities and predicted white dwarf planet distribution with the EDR3 white dwarf catalogue \citep{gentilefusillo_catalogue_2021} as described in \S \ref{methods-number} and \S\ref{results-number}, predictions are made for the number of planet detections around white dwarfs with \Gaias. The implications of these predictions and how they can be used in the future in combination with \Gaia planet detections is outlined in \S\ref{discussion}, alongside a discussion of their validity. Finally, we conclude in  \S\ref{conclusion}.

\section{Methods} \label{methods}

The aim is to predict the population of planets around white dwarfs, as seen by \Gaias, that evolve from the population of planets seen around main-sequence stars. We consider the fundamental dynamical effects on the orbit of stellar mass loss and tides. Further complications are discussed in \S\ref{discussion} and comparison between the predictions in this paper and \Gaia detections will determine the importance of these additional processes, including photoevaporation and dynamical scattering.  

In order to predict the number of planets \Gaia will find around white dwarfs that evolved from the main-sequence by tides and mass loss, we consider three main steps: 

\begin{enumerate}
    \item Calculation of \textit{Gaia} detection probability as a function of planet mass and semi-major axis.
    \item Prediction of a post-main sequence planet distribution as a function of mass and semi-major axis.
    \item Convolution of the detection probability and post-main sequence planet distribution as functions of mass and semi-major axis and summation over this convolution to determine how many planets \textit{Gaia} should find around white dwarfs.
\end{enumerate} The method for each of these steps is described in this section. 

\subsection{\textit{Gaia} detection probabilities as a function of mass and semi-major axis}\label{methods-probability}
\begin{figure*}
    \centering
    \includegraphics[width=1\textwidth]{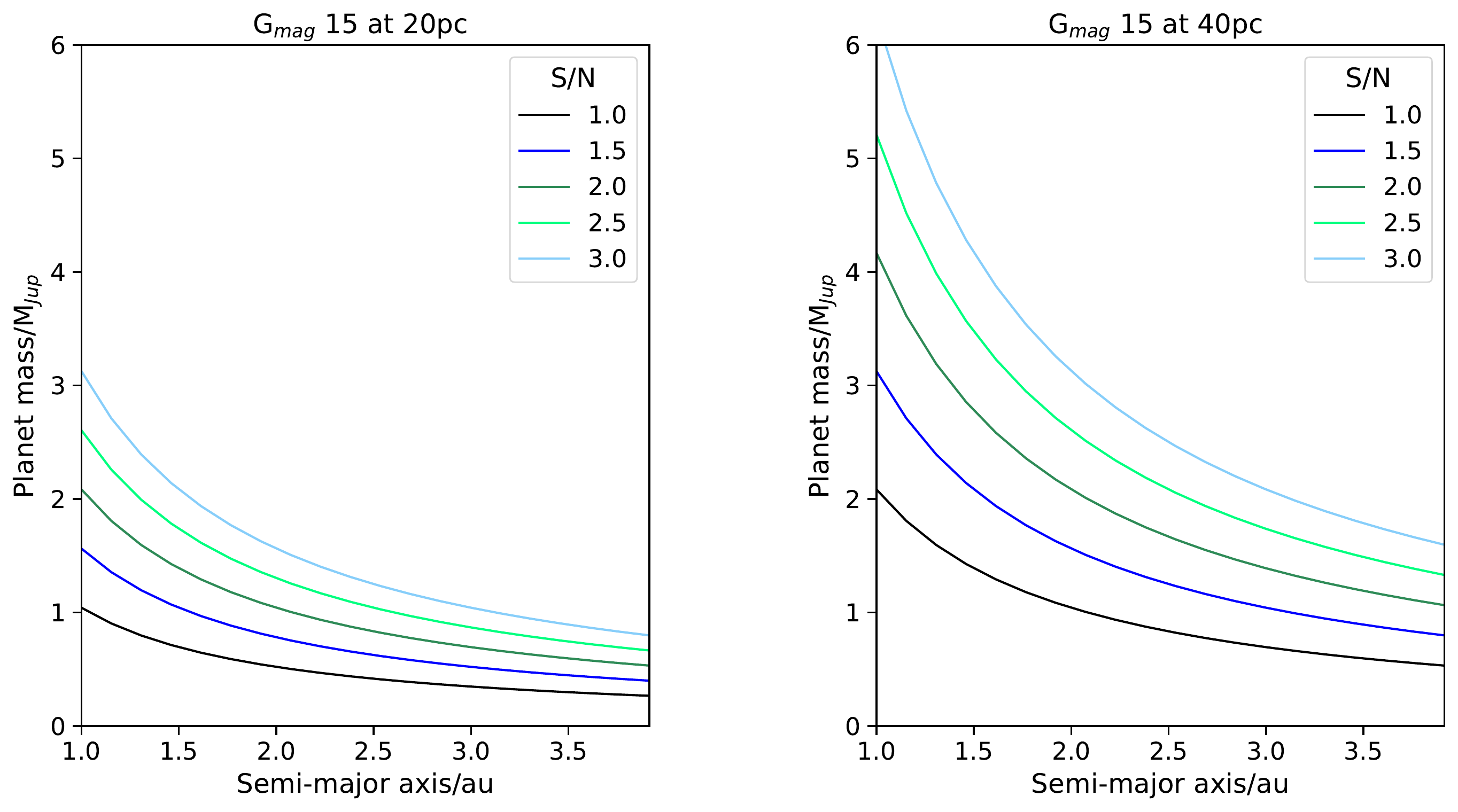}
    \caption{The astrometric signal for a planet with given properties (mass and semi-major axis) compared to the astrometric noise for \Gaia measurements of a 15th mag white dwarf at 20\,pc (left) or 40\,pc (right). The semi-major axis range spans positions from the innermost initial position in our simulations to the maximum semi-major axis detectable by \textit{Gaia} due to its ten year mission length (see \S \ref{methods-probability}). Planets with S/N>3 have an average detection probability of 0.93, whilst planets with S/N<1 have average detection probability of 0.14.}
    \label{fig:40pc_contours}
\end{figure*}
\begin{figure}
    \centering
    \includegraphics[width=1\columnwidth]{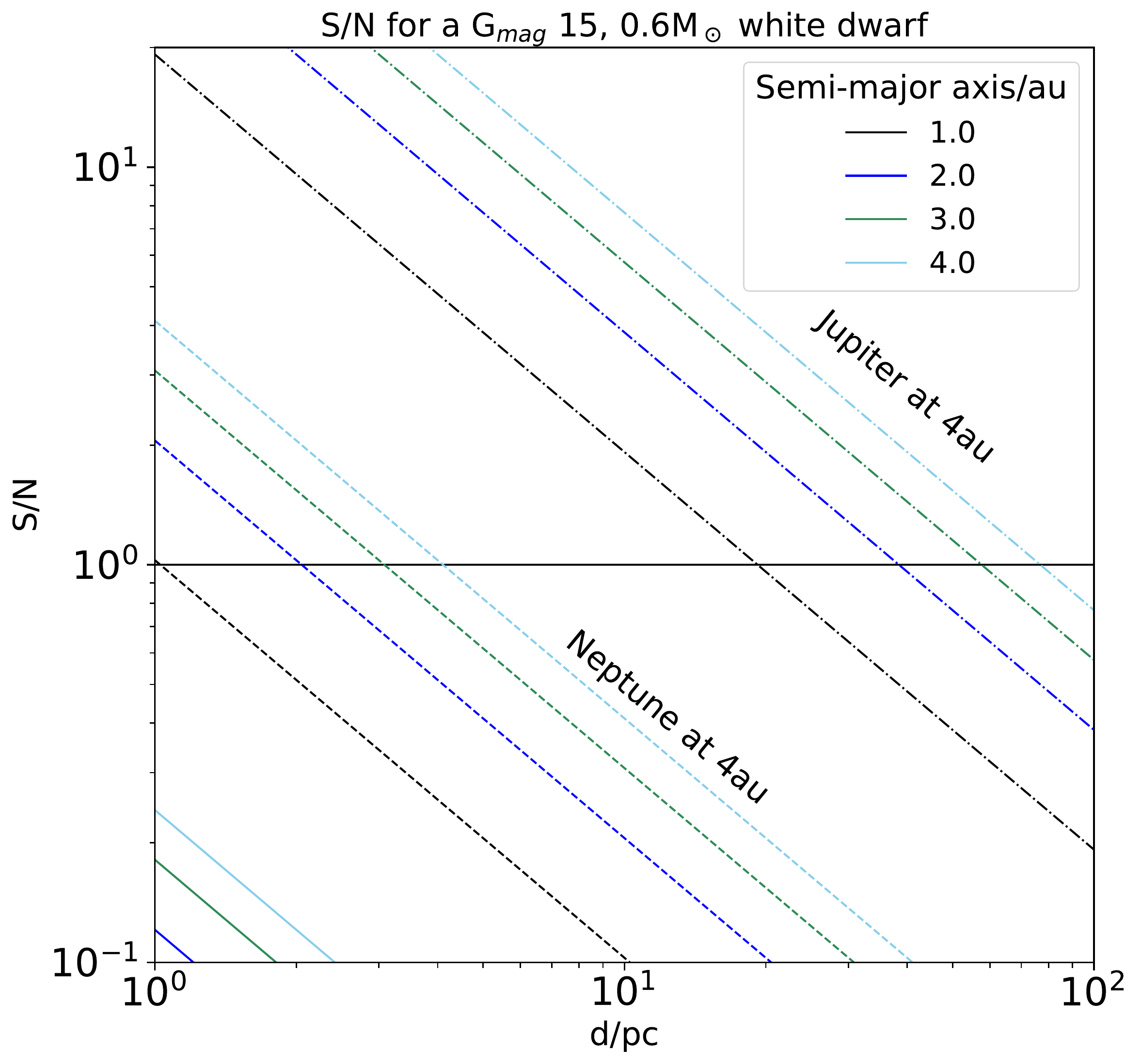}
     \caption{The astrometric signal of a Jupiter or Neptune mass planet on a circular orbit at 1-4\,au, compared to the astrometric noise in a \Gaia measurement for a star at a given distance. The maximum semi-major axis corresponds to the maximum detectable by \textit{Gaia} on a ten year mission observing a $0.6M_\odot$, $G_{\tm{mag}}=15$ white dwarf. Dot-dash lines are for a Jupiter mass planet, dash lines are for a Neptune mass planet. A solid horizontal line is for a $S/N=1$ below which the detection probability drops to zero. Jupiter mass planets are only detectable out to 30\,pc. Neptune mass planets are not detectable, because their signal to noise drops below one at 3\,pc and there are no white dwarfs within 5\,pc \citep{hollands_gaia_2018,gentilefusillo_catalogue_2021}.}
    \label{fig:loglogplanets}
\end{figure}

In this paper, a signal to noise (S/N) criterion, relating astrometric signal of a planet to the single along scan accuracy per field of view of a \Gaia measurement, is used to link planet mass and semi-major axis to detection probability. This is the simplest approach to reliably understand which planets have a strong enough signal to be detected by \Gaia and is commonly used as indicator when comparing different planet detection algorithms \citep[e.g. in][]{casertano_double-blind_2008,ranalli_astrometry_2018,perryman_astrometric_2014}. Other subtler signatures, such as RUWE, can also currently be used as an indicator of binarity \citep{belokurov_unresolved_2020,penoyre_astrometric_2022}.

The astrometric signature, $\alpha$, is approximated by the ratio of planet and stellar masses ($M_{\tm{pl}}$ and $M_\star$) in solar masses, their separation, $a$, and the distance to the star, $d$, as shown in Equation \ref{eq:signature}.
\begin{equation}
    \alpha=\left(\frac{M_{\tm{pl}}}{M_{\star}}\right)\left(\frac{a}{1\,\tm{au}}\right ) \left(\frac{d}{1\,\tm{pc}}\right)^{-1}as
\label{eq:signature}\end{equation}
More massive planets or those further from the star have larger signatures, because they displace the barycentre position further from the centre of the star. The signature is inversely proportional to the distance to the system, because \textit{Gaia} measures angular positions of stars and the angle subtended by the distance between the star and the system's barycentre will decrease with increasing distance to the system. We consider it sufficient to use an average white dwarf mass $M_\star=0.6\,M_\odot$ \citep{hollands_gaia_2018} in this work, when calculating $\alpha$ , because the difference in white dwarf masses is small and makes little difference to $\alpha$ in comparison to the large change in astrometric noise with white dwarf magnitude. Additionally, there are several sources of error on the mass values in the catalogue, as discussed in \S\ref{discussion-mass-cut}.

Signal can be maximised by looking for planets at large semi-major axes, \textit{a}. However, the ability of \textit{Gaia} to accurately recover a companion's orbital parameters is limited by mission length. For planets with orbital periods longer than the expected mission length following extensions (ten years), the accuracy of the  planet orbital parameters which are recovered decreases \citep{ranalli_astrometry_2018}. Therefore in this work, a period of ten years has been taken as the cut off for reliable planet detections. This corresponds to $a=4.64$\,au for a planet around a $1M_\odot$ star or $a=3.91$\,au for a planet around a $0.6M_\odot$ white dwarf. 

The noise on \textit{Gaia}'s astrometric measurement is the single along-scan accuracy per field of view, $\sigma_{\tm{fov}}$ \citep{perryman_astrometric_2014}. Importantly, $\sigma_{\tm{fov}}$ is a function of a star's magnitude. In this work the S/N ratio will be defined as \begin{equation}S/N=\frac{\alpha}{\sigma_{\tm{fov}}}. \label{eq:s_to_n} \end{equation}
\textit{Gaia} errors for future data releases are publicly available as sky-averaged parallax accuracy $\sigma_{\varpi}$ not as $\sigma_{\tm{fov}}$. $\sigma_{\varpi}$ can be converted to $\sigma_{\tm{fov}}$ by considering the sky-averaged number of field crossings per star ($150\pm50$ for a ten year mission \citep{ranalli_astrometry_2018}), the geometrical factor linking the sky-averaged parallax accuracy with the error per field crossing (2.15) and a science contingency margin (1.1 for EDR3 onwards), which covers any additional errors in the modelling process \citep{perryman_astrometric_2014}.\begin{equation}
    \sigma_{\tm{fov}}=\frac{\sqrt{150}\sigma_{\varpi}}{1.1\times2.15}\approx5.18\sigma_{\varpi} \label{eq:error conversion}
\end{equation}
This paper uses the predicted errors for Data Release 5, because this data release will have the longest timespan for observations so can detect planets with periods up to ten years. These error estimates are based on \textit{Gaia} EDR3 and are the most up-to-date at the time of press.
The errors are described as: \protect\footnotemark   \footnotetext{Taken from \textit{Gaia} Mission Science Performance for Data Release 5 \url{https://www.cosmos.esa.int/web/gaia/science-performance}. } \begin{equation}\label{eq:errors}
   \sigma_{\varpi}[\mu\,as]=0.527(40+80z+30z^2)^{\frac{1}{2}},\end{equation}
where $z$ is given by 
\begin{equation}\label{eq:z}
z=MAX[10^{0.4(13-15)},10^{0.4(G-15)}]
\end{equation} where \textit{G} is the broad-band, white-light, \Gaia magnitude. This S/N criterion describes the accuracy of a measurement. Visibility periods are used as a measure of reliability \citep{LL:LL-124}: the more visibility periods a source has, the less the astrometric solution will be affected by a bad measurement. This is used to filter sources before fitting the astrometric solution \citep[e.g. visibility\_periods\_used >11][]{arenou_gaia_2022}. Once DR5 is available, visibility periods for each source can be used to assess the reliability of the astrometry before looking for companions. 

S/N can be linked to detection probability. \citet{ranalli_astrometry_2018} used Markov chain Monte Carlo methods and three information criteria to determine detection probabilities over a grid of S/N and orbital period. Planets with S/N>3 have an average detection probability (across periods less than 10 years) of 0.93, whilst planets with S/N<0.7 cannot be detected. In this work, $p_{ljk}(S/N_{ljk},P_j)$ denotes the detection probability of a planet with semi-major axis in bin $j$ with period $P_j$, mass in bin $k$ around white dwarf $l$ and were calculated from the \citet{ranalli_astrometry_2018} detection probabilities.

\subsection{Prediction of the distribution of planets which survive tidal evolution and stellar mass loss to reach the white dwarf phase} \label{methods-distribution}

The distribution of close-in planets around main-sequence stars is well known from radial velocity and transit observations. Some of these observations, alongside direct imaging surveys, now reach orbital periods of years, as probed by \Gaias, but the planets detected by \Gaia around main-sequence stars will always provide the most reliable probe of the planetary population of interest on the main-sequence.

Radial velocity \citep{bryan_statistics_2016}, transit \citep{zhu_exoplanet_2021,santerne_sophie_2016} and direct imaging \citep{biller_gemininici_2013} measurements suggest the frequency of giant planets decreases at distances greater than 2-3\,au from the star, such that a broken power law is the most appropriate model for the giant planet occurrence rate around main-sequence stars \citep{fernandes_hints_2019,fulton_california_2021}. 
In this work, we use the EPOS, three parameter fit (symmetric) broken power law of \citet{fernandes_hints_2019}\footnote{The errors in \citet{fernandes_hints_2019} are asymmetric, but this work uses the average of the upper and lower bounds as an estimate of the size of the error.}. \begin{equation}
    \frac{\tm{d}^2N}{\tm{d\,log}P \:\tm{d\,log}M}=c_0\left(\frac{P}{P_{\tm{break}}}\right)^{p_1}\left(\frac{M}{10M_\oplus}\right)^{m_1}
\end{equation} where $c_0=0.84\pm0.17$, $m_1=-0.45\pm0.05$, $p_{\tm{break}}=1581\pm643$ days and \begin{equation}
    p_1= \begin{cases} 0.65 \pm 0.17 & P<P_{\tm{break}} \\ -0.65 \pm 0.17 & P>P_{\tm{break}} \end{cases}.
\end{equation} This broken power law describes the number of planets per log semi-major axis and log mass bin, $\frac{\tm{d}^2N}{\tm{d\,log}P \:\tm{d\,log}M}$ and models the distribution of planets around main-sequence stars accounting for observation biases. This planet occurrence rate predicts $4.9\pm0.7\%$ of stars have planets with masses $1-13\,M_J$ with semi-major axes <\,20\,au. Non-detections are considered by weighting the number of detected planets in a mass and semi-major axis bin by the inverse of the survey completeness. The power law of \citet{fernandes_hints_2019} was used because it reproduced \citet{cumming_keck_2008} earlier results, gave results comparable to direct imaging observations at large semi-major axes and drew on multiple groups of observations (the latest Kepler data release and radial velocity results from \citet{mayor_harps_2011}). 

To predict the change in planetary orbits from the main-sequence to the white dwarf phase, the orbital evolution of planets along the AGB was simulated, following the prescription presented in \citet{mustill_foretellings_2012}. This model breaks the effects of stellar evolution on planets into three key components: expansion of the planetary system due to mass loss, inward movement of planets due to tidal effects and engulfment of planets which are too close to the star as the star expands. Tidal forces were modelled as viscous dissipation of the equilibrium tide \citep{zahn_tidal_1977}. The model calculates the semi-major axis of the orbits of isolated planets and the stellar envelope as a star evolves along the AGB, following the models of \citet{vassiliadis_wood_1993}. Any planet which moves inside the stellar envelope at a given timestep is removed. The limitations of this model are discussed in \S\ref{dis-model}.

Extending the masses and initial semi-major axes of \citet{mustill_foretellings_2012}, we simulated 13,051 planets around a $1M_\odot$ progenitor and 12,121 planets around a $1.5M_\odot$ progenitor\footnote{Fewer simulations were needed for the higher mass progenitor because the larger expansion of the planetary system moved most of the sample outside the \Gaia detection range.}. For simplicity, all planets were assumed to be in single planet systems on circular orbits (see \S \ref{dis-other}). The planet masses sampled 31 possible values evenly distributed in log space between Earth and 13 Jupiter masses. Initial semi-major axes ranged from 1-10\,au. All planets at semi-major axes smaller than this would be engulfed by their host star and those beyond this would be far beyond the \Gaia detection region so were not considered. Two groups of simulations were carried out. Simulation A was a coarse spacing of 0.03\,au to enable sampling of a large range of semi-major axis space. Simulation B and C were a fine spacing of 0.001\,au spanning $\pm 0.12$\,au ($1\,M_\odot$) and $\pm 0.09$\,au ($1.5\,M_\odot$) from the initial position of the outermost engulfed planet (determined from Simulation A) at each mass. This fine sampling was to investigate behaviour near the initial semi-major axis of the outermost engulfed planet, because in this region there is large variation in the strength of the tidal forces \citep[see][]{mustill_foretellings_2012}. The parameters for these simulations are summarised in Table \ref{tab:sim_spacing}. The final masses of the white dwarfs (needed for estimating the importance of tidal effects) were $0.5702M_\odot$ and $0.6014M_\odot$ for the $1M_\odot$ and $1.5M_\odot$ progenitors respectively.
\begin{table*}
    \centering
    \begin{tabular}{|c|c|c|c|c|}
  
        Simulation & Progenitor mass/$M_\odot$ & Spacing/au & Planet mass/$M_J$ & Semi-major axis range/au
        
        \\\hline
        A& 1.0, 1.5 & 0.03 & $3.1\times10^{-4}-13$& 1-10
         \\ \hline B & 1.0 & 0.001 & $3.1\times10^{-4}-13$ & inner position of outermost engulfed planet $\pm0.12$
         \\ \hline C & 1.5 & 0.001 & $3.1\times10^{-4}-13$ & inner position of outermost engulfed planet $\pm0.09$
        
    \end{tabular}
    \caption{Semi-major axis ranges for simulated planets for each planet mass. 31 planet masses were chosen equally spaced in log space between $M_\oplus$ and $13\,M_J$. For each planet mass, the above semi-major axes arrays were used for the initial conditions of the simulations. All planets were assumed to be in single planet systems on circular orbits (see \S \ref{dis-other}). For discussion of choices of semi-major axis spacing see \S \ref{methods-distribution}.}
    \label{tab:sim_spacing}
\end{table*}

To combine the main-sequence planet occurrence rate with the results of our simulations, mass and semi-major axis space were split into bins. There were 31 bins equally spaced in log$M$ from $1M_\oplus$ to $13\,M_J$ and the semi-major axis bins have variable width corresponding to equispaced period bins from \citep{ranalli_astrometry_2018} (average width $0.21$\,au). This corresponded to the range of initial positions in our simulation. We assumed the planets were around a 0.6$M_\odot$ white dwarf to transform the main-sequence distribution from $\frac{\tm{d}^2N}{\tm{d\,log}P \:\tm{d\,log}M}$ to $\frac{\tm{d}^2N}{\tm{d\,log}a \:\tm{d\,log}M}$ where $N$ is the number of planets per star, $a$ is the semi-major axis, $P$ is the period and $M$ is the planet mass. All logarithms are to base 10 so $\tm{d\,log}a=\frac{\tm{d}a}{a}\tm{log}_{10}(e)$ was used to transform from $\frac{\tm{d}^2N}{\tm{d\,log}a \:\tm{d\,log}M}$ to $\frac{\tm{d}^2N}{\tm{d}a \:\tm{d\,log}M}$ where d$a$ was the corresponding bin width and $a$ was the midpoint of the bin. In the following equations, the index \textit{i} represents the initial semi-major axis bin of a planet, \textit{j} represents the final semi-major axis bin and \textit{k} represents the mass bin. Planetary mass loss was not modelled so a planet's mass bin does not change during the simulation. All summations are written explicitly.

The predicted occurrence rate of planets around white dwarfs in a given semi-major axis, mass range, $\frac{\tm{d}^2N_{\tm{WD}}}{\tm{d}a\:\tm{d\,log}M}$ is calculated by \begin{equation}
  \left(\frac{\tm{d}^2N_{\tm{WD}}}{\tm{d}a\:\tm{d\,log}M}\right)_{jk} = \Sigma_i R_{ijk}C_{ik}. 
\label{eq:pred_dist}\end{equation} $R_{ijk}$ is the number of planets of mass labelled by $k$ in the simulation that started in the $i^{th}$ semi-major axis bin, but end the simulations in the $j^{\tm{th}}$ semi-major axis bin post-main sequence. $C_{ik}$ is the weight of each simulated planet, which is the value of the main-sequence planet distribution from \citet{fernandes_hints_2019} for the centre of the bin, divided by the number of simulated planets initially in the $i^{th}$ bin, $B_i$.  \begin{equation}
    C_{ik}=\left(\frac{\tm{d}^2N_{\tm{MS}}}{\tm{d}a\:\tm{d\,log}M}\right)_{ik}\frac{1}{B_i}
\label{eq:count_weight}\end{equation}  

\subsection{Predicted numbers of planet detections}\label{methods-number}
\subsubsection{Using the EDR3 catalogue}
To predict the number of planets \Gaia will detect around white dwarfs requires the calculation of the parameter space available for planet detection around known white dwarfs. The white dwarf sample, containing mass estimates, \Gaia broad-band magnitudes ($G_{\tm{mag}}$) and parallaxes was obtained from the  \citet{gentilefusillo_catalogue_2021} catalogue from \textit{Gaia} early Data Release 3, which improved on their previous catalogue from Data Release 2 \citep{gentilefusillo_gaia_2019}. In \citet{gentilefusillo_catalogue_2021}, each object in the catalogue was assigned a probability of being a white dwarf $P_{\tm{WD}}$ based on a series of absolute magnitude and colour critera. In \citet{gentilefusillo_catalogue_2021}, applying the criteria $P_{\tm{WD}}>0.75$ recovered 359,000 high confidence white dwarf candidates with only 1$\%$ of contaminant objects. Therefore in this work, we applied three filters to the catalogue to obtain a catalogue of white dwarfs: \begin{itemize}
    \item $P_{\tm{WD}}>0.75$ 
    \item $G_{\tm{mag}}<20.7$ - remove objects too faint to be detected by \textit{Gaia}.
    \item $M< 0.663\,M_\odot$ - remove white dwarfs with progenitor masses $>1.5\,M_\odot$.
\end{itemize} After applying these cuts, the sample is reduced to 282,718 white dwarfs out to distances of 16,000\,pc. Planet detections around the furthest away white dwarfs in this sample will not be possible due to the high noise and small astrometric signature. For a $G_{\rm mag}=20.7$ white dwarf, there will be a 50\% relative uncertainty on position at approximately 800\,pc. Although an \textit{a priori} distance cut was not applied, they will not have contributed to the detection probability value as their S/N was so small. 

White dwarfs with high progenitor masses were excluded from the catalogue, because they lose a greater proportion of mass as they evolve off the main-sequence \citep{cummings_white_2018} and the surviving planet positions are beyond the maximum semi-major axis detectable by \textit{Gaia} for a ten year mission. This will be discussed further in \S \ref{results-planet-distribution} (Figure \ref{fig:final_distribution1p5}), where the chosen cut-off is justified based on simulations that indicate that no planets directly evolved from the main-sequence will be detected around stars within initial masses higher than $M_i>1.5M_\odot$. The final mass cut off was determined using the \citet{cummings_white_2018} MIST-based initial-final mass relation based on stellar clusters. This gave $M_f=0.609\pm0.054M_\odot$ for a $1.5M_\odot$ progenitor, so an upper limit of $0.663\,M_\odot$ was adopted. White dwarfs in the \citet{gentilefusillo_catalogue_2021} catalogue with unknown masses were assumed to have a mass equal to $0.6\,M_\odot$ \citep[the average white dwarf mass][]{hollands_gaia_2018}.

\subsubsection{Combining the predicted planet distribution with \Gaia detection probabilities}
For each white dwarf in the filtered catalogue, labelled by the index $l$, the probability of detecting a planet as a function of planet mass, $j$ and planet semi-major axis, $k$, $p_{ljk}$ was calculated as described in \S \ref{methods-probability}. The semi-major axis bins were determined from the period bins in \citet{ranalli_astrometry_2018}. The mass bins were unchanged. $\left(\frac{\tm{d}^2N_{\tm{WD}}}{\tm{d}a\:\tm{d\,log}M}\right)_{jk}$ calculated in \S\ref{methods-distribution} was recalculated to match these bins for the following calculations.

The distribution of detected planets $\frac{\tm{d}^2N_{\tm{det}}}{\tm{d}a\:\tm{d\,log}M}$ corresponds to \begin{equation}
    \left(\frac{\tm{d}^2N_{\tm{det}}}{\tm{d}a\:\tm{d\,log}M}\right)_{jk}=\Sigma_lp_{ljk}\left(\frac{\tm{d}^2N_{\tm{WD}}}{\tm{d}a\:\tm{d\,log}M}\right)_{jk}.
\label{eq:det_dist}\end{equation}
The number of detected planets is
\begin{equation}
    N_{\tm{pl}}=\Sigma_{jk}\left(\frac{\tm{d}^2N_{\tm{det}}}{\tm{d}a\:\tm{d\,log}M}\right)_{jk}(\tm{d}a)_j(\tm{d\,log}M)_k.
\label{eq:npl}\end{equation} 

\subsubsection{Error Analysis}\label{methods-error}
The error on the number of planets was calculated by appropriately propagating the errors on each of the quantities in equations \ref{eq:pred_dist}-\ref{eq:npl}. The error on $\left(\frac{\tm{d}^2N_{\tm{MS}}}{\tm{d}a\:\tm{d\,log}M}\right)_{ik}$ came from the average of the upper and lower bounds on these values from \citet{fernandes_hints_2019}. The error on $R_{ijk}$ (the number of planets of mass labelled by $k$ in the simulation that moved from the $i^{\tm{th}}$ to the $j^{\tm{th}}$ semi-major axis bin in the simulations)  was assumed to be Poissonian $\sigma(R_{ijk})=\sqrt{R_{ijk}}$. The dominant error in the detection probabilities is from the error in the sky-averaged number of field crossings per star ($150\pm50$)  \citep{ranalli_astrometry_2018}. The resulting error in the detection probabilities was combined fractionally in quadrature with the error on $\left(\frac{\tm{d}^2N_{\tm{WD}}}{\tm{d}a\:\tm{d\,log}M}\right)_{ik}$ to obtain the error on $\left(\frac{\tm{d}^2N_{\tm{det}}}{\tm{d}a\:\tm{d\,log}M}\right)_{ik}$.
Another possible source of uncertainty is the exclusion white dwarfs with $M>0.663\,M_\odot$. This is difficult to quantify so was not included in our error calculations but is discussed in \S \ref{discussion-mass-cut}.

\section{Results} \label{results}
\subsection{What is the probability of \Gaia detecting a planet as a function of mass and semi-major axis?}
\label{results-probability}

Only planets with masses greater than or equal to Jupiter will be detectable by \textit{Gaia} around white dwarfs. The astrometric signal of a planet on a given orbit with a given mass is calculated according to Equation \ref{eq:signature} and the noise is calculated by \ref{eq:error conversion}.  Figure \ref{fig:loglogplanets} indicates that Neptune mass planets have $S/N>1$ up to 3\,pc and beyond this only Jupiter mass planets have $S/N>1$. Since the closest white dwarf, 40 Eri B, is at 5.04\,pc  \citep{hollands_gaia_2018,gentilefusillo_catalogue_2021}, this suggests no Neptune mass planets will be discovered around white dwarfs by \textit{Gaia}. As shown in Figure \ref{fig:40pc_contours}, even Jupiter mass planets have a low S/N and correspondingly a low detection probability at small semi-major axes
These calculations were done for a bright white dwarf with $G_{\tm{mag}}=15$. For fainter white dwarfs, the astronometric noise would be larger, decreasing the S/N of a given planet mass and semi-major axis.

\begin{figure*}
    \centering
    \includegraphics[width=2\columnwidth]{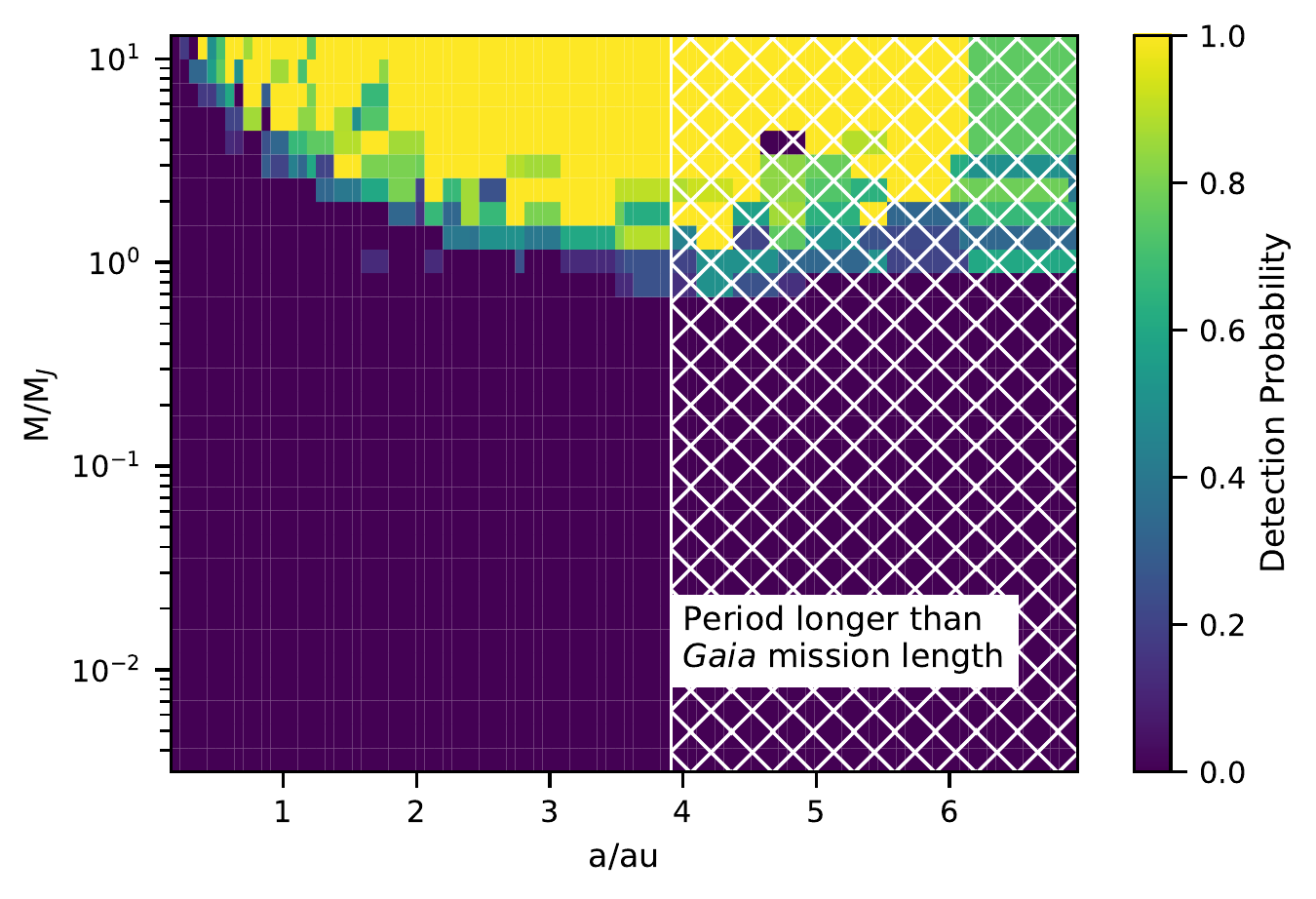}
    \caption{
   Detection probability, $p_{ljk}$ as a function of mass and semi-major axis for an example $G_{\tm{mag}}=15$ white dwarf at 30\,pc. Here detection probability refers to the likelihood of detecting a given planet if it exists around the white dwarf and does not account for $\frac{\tm{d}^2N_{\tm{WD}}}{\tm{d}a\:\tm{d\,log}M}$.The hatched region indicates semi-major axes which have non-zero detection probabilities, but periods longer than the mission length. The dark square around 4.5\,au and 3$\,M_J$ results from the stochastic method to determine the detection probabilities in \citep{ranalli_astrometry_2018}. Almost all planets above $3\,M_J$ at a few au will be detected, whilst almost no planets less massive than Jupiter will be detected. }
    \label{fig:detect_prob}
\end{figure*}

Figure \ref{fig:detect_prob} shows an example map illustrating detection probability as a function of mass and semi-major axis generated for a $G_{\tm{mag}}=15$ white dwarf. It clearly demonstrates that almost all planets less massive than Jupiter will not be detected by Gaia. There is a significant region of parameter space where planets with orbital periods longer than the mission length could be detected, shown by the hashed region in Figure \ref{fig:detect_prob}, however the recovery of orbital parameters in this region is unreliable. For clarity, planet detections in this region are not included in the predicted number of planets \Gaia will detect (\S\ref{results-number}).

\subsection{What is the planet distribution around white dwarfs?}\label{results-planet-distribution}
\begin{figure*}
    \centering
    \includegraphics[width=1\textwidth]{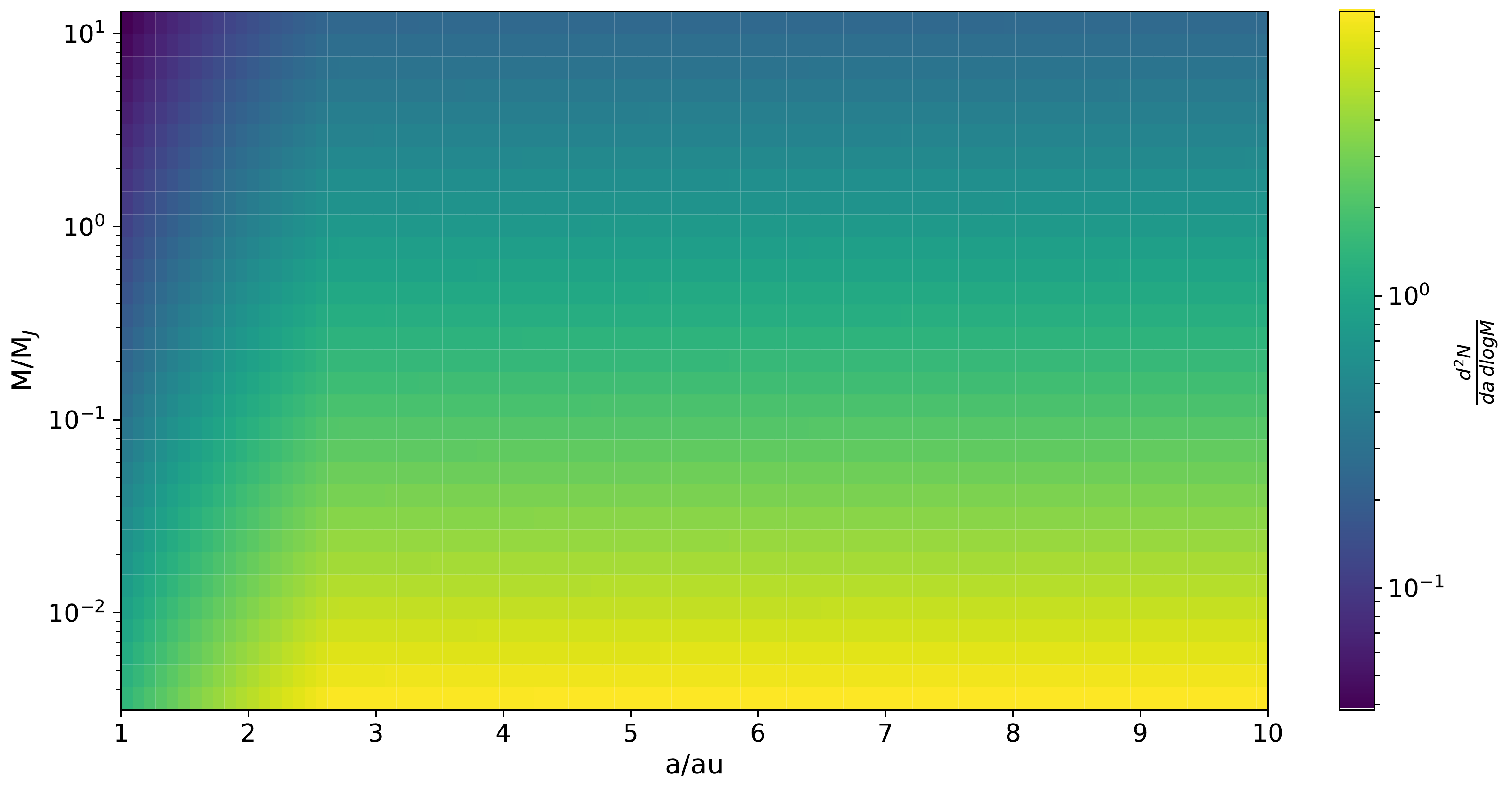}
    \caption{Number of planets per semi-major axis and log M bin  ($\frac{\tm{d}^2N_{\tm{MS}}}{\tm{d}a\:\tm{d\,log}M}$) for main-sequence stars from \citet{fernandes_hints_2019}. Lower mass planets are more common.}
    \label{fig:initial_dist}
    \end{figure*}

\begin{figure*}
\begin{subfigure}[b]{1\textwidth}
    \centering
    \includegraphics[width=1\textwidth]{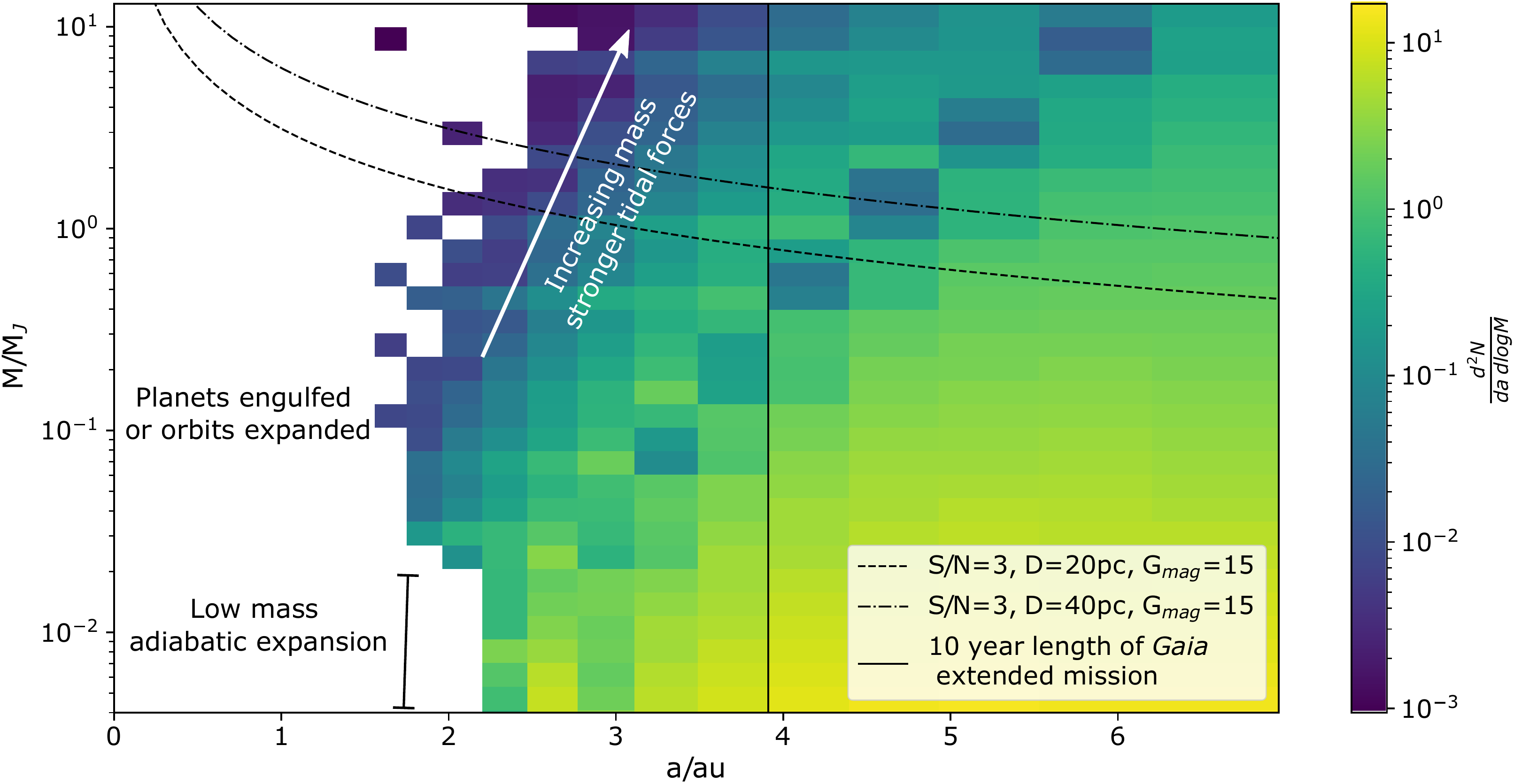}
    \caption{1\,$M_\odot$ progenitor. A small proportion of planets survive inside the detection region to the white dwarf phase (see \S \ref{results-planet-distribution}).}
    \label{fig:final_distribution1p0}
\end{subfigure}
\begin{subfigure}[b]{1\textwidth}
    \centering
    \includegraphics[width=1\textwidth]{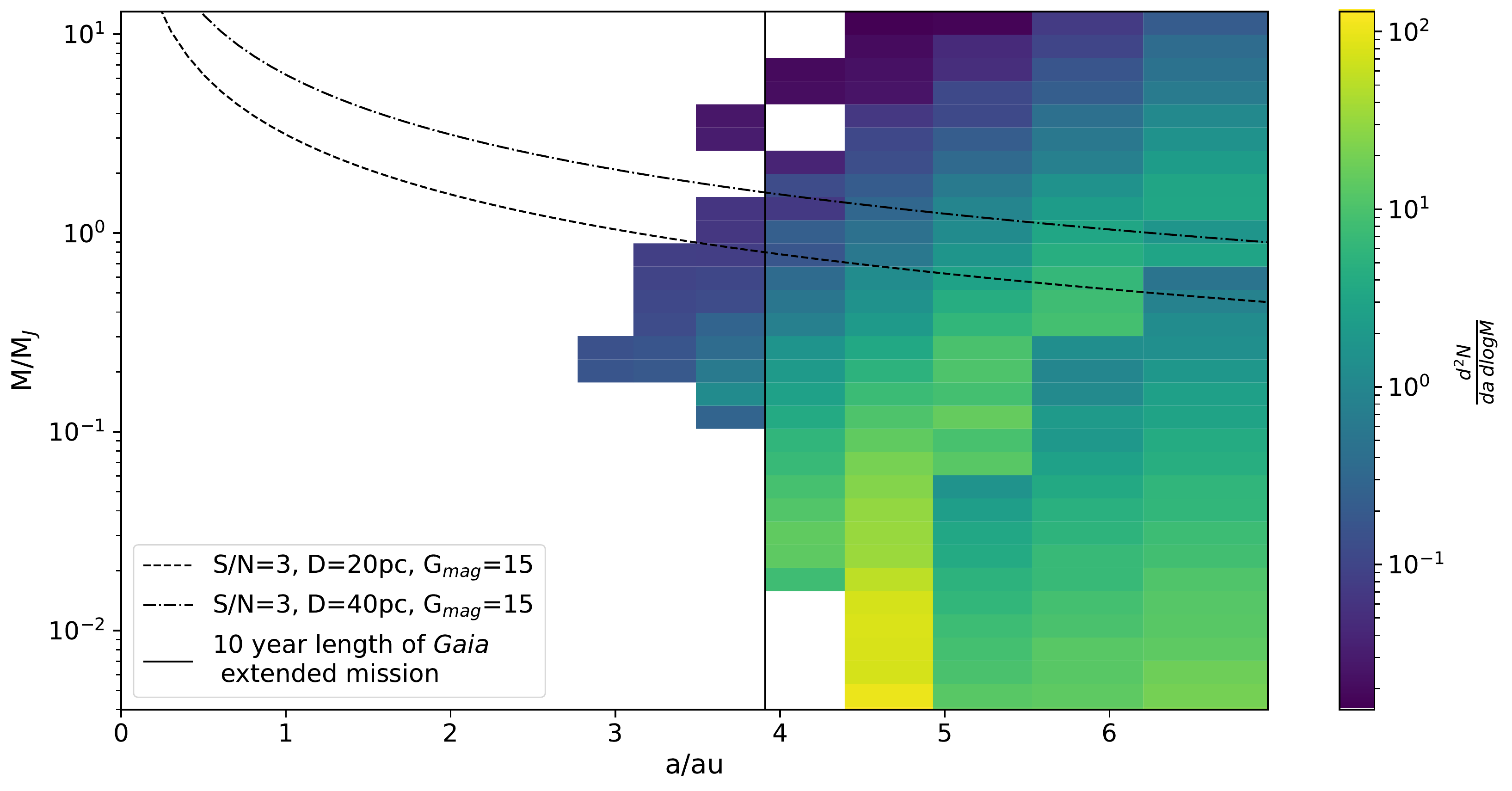}
    \caption{$1.5\,M_\odot$ progenitor. Stronger tidal effects and greater mass loss means fewer planets survive inside the detection region compared to the $1M_\odot$ progenitor.}
    \label{fig:final_distribution1p5}
\end{subfigure}
\caption{Predicted number of planets per semi-major axis and log M bin  ($\frac{\tm{d}^2N_{\tm{WD}}}{\tm{d}a\:\tm{d\,log}M}$) around a white dwarf with a 1$M_\odot$ and 1.5$M_\odot$  progenitor. This is based on the evolution of an initial distribution around main-sequence stars (\citep{fernandes_hints_2019} see Fig \ref{fig:initial_dist}) taking into account tidal evolution, mass loss and engulfement by the AGB star (see \S \ref{methods-distribution} for further details). The semi-major axis bins have variable width corresponding to equispaced period bins from \citep{ranalli_astrometry_2018} and  have average width 0.21\,au. The mass bins are $0.12$ wide in $\tm{log}(M/M_J)$. The solid black line denotes the semi-major axis with a period equal to the length of the \Gaia extended mission for a planet orbiting a $0.6M_\odot$ white dwarf. The orbital parameters of planets at semi-major axes greater than this cannot be recovered accurately. The dashed and dot-dashed lines represent the detection cut off for a signal-to-noise of three for a $G_{\tm{mag}}=15$ white dwarf at 20\,pc and 40\,pc respectively. Trends discussed in \S \ref{results-planet-distribution} have been labelled on Figure \ref{fig:final_distribution1p0}.}
\end{figure*}

Predictions for the population of planets surviving to the white dwarf phase are shown in Figure \ref{fig:final_distribution1p0} for a $1\,M_\odot$ progenitor. These should be compared to the initial population of planets on the main-sequence \citet{fernandes_hints_2019} shown in Figure \ref{fig:initial_dist} as $\frac{\tm{d}^2N_{\tm{MS}}}{\tm{d}a\:\tm{d\,log}M}$.
Over the mass and semi-major axis range shown here the main-sequence distribution gives a planet occurrence rate of $0.61\pm0.01$ planets per star. This is to be contrasted with the population surviving to the white dwarf phase, where only $14\%$ of planets initially within 3.91\,au (the detection limit due to the length of \Gaia mission) of $1\,M_\odot$ progenitor remain.
The number of planets per star in the \Gaia detection region is lower than the main-sequence, because planets are engulfed or move out of this region as their orbits expand due to stellar mass loss post-main sequence. This is evident from the white regions in Figures \ref{fig:final_distribution1p0} where $\frac{\tm{d}^2N_{\tm{WD}}}{\tm{d}a\:\tm{d\,log}M}$ is zero. Even fewer planets ($0.1\%$) survive in the detection region around a $1.5\,M_\odot$ progenitor, because larger mass progenitors have stronger tidal forces and lose a greater proportion of their mass. The majority of the region inside 3.91\,au in Figure \ref{fig:final_distribution1p5} is white. This means it is very unlikely that planets will be detected around white dwarfs with $M>1.5\,M_\odot$ progenitors. This is the basis for the mass cut off applied to the EDR3 catalogue (\S\ref{methods-number}), excluding white dwarfs with higher progenitor masses. 

Tidal evolution of gas giants leads many planets to migrate inwards: some migrate so far they are engulfed by the stellar envelope. This means that only planets with initial positions sufficiently far out survive to the white dwarf phase. This distance increases with planet mass, as higher mass planets feel a stronger tidal pull from the star. This is demonstrated by the diagonal trend of the boundary between zero and non-zero values of $\frac{\tm{d}^2N_{\tm{WD}}}{\tm{d}a\:\tm{d\,log}M}$ in Figures \ref{fig:final_distribution1p0} and \ref{fig:final_distribution1p5}. Planets beyond a certain distance are too far away to be affected by tides (see Appendix \ref{app-tides}). Planets below $10^{-2}\,M_J$ are not massive enough to be affected by tidal forces and their orbits undergo a purely adiabatic expansion, as shown by the vertical portion of the boundary in Figure \ref{fig:final_distribution1p0}.

\subsection{Predictions for the number of planets \Gaia will detect around white dwarfs}\label{results-number}
\begin{figure*}
    \centering
    \includegraphics[width=2\columnwidth]{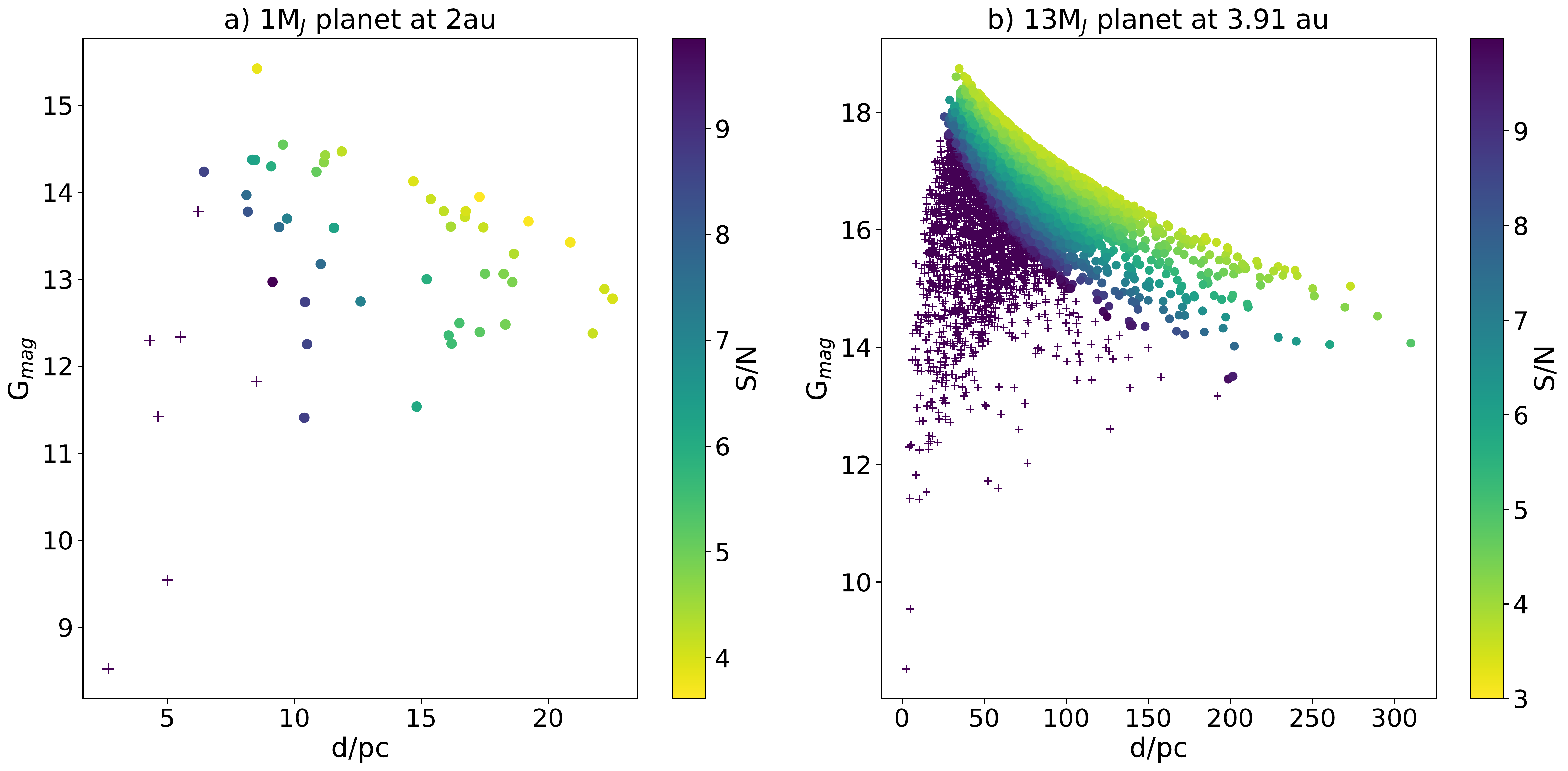}
    \caption{The properties of the white dwarfs ($G_{\tm{mag}}$ and distance) for whom detection of a $1\,M_J$ at 2\,au (weak signal) and a $13\,M_J$ at 3.9\,au is guaranteed if that planet exists around those white dwarfs (detection probability = 1). Candidates with $S/N>10$ are denoted by the purple +. There are an far fewer candidates, 3 (see Appendix \ref{app-table}), for detecting lower mass ($\sim M_J$) close in (<2\,au) compared to 3770 for a $13\,M_J$ planet at 3.91\,au. Also note the difference in semi-major axis and $G_{\tm{mag}}$ scales between the two plots.}
    \label{fig:wd_dist}
\end{figure*}

The best white dwarf candidates for finding planets are bright (smaller noise) and close (larger signal). $50\pm10\%$ of planet detections occur around white dwarfs brighter than $G_{\tm{mag}}=15$, even though they comprise just 441 out of the 282,718 stars in the catalogue. There are far fewer white dwarfs where detection of a close in Jupiter mass planet is likely compared to a $13\,M_J$ planet at large semi-major axes (52 compared to 5519) as demonstrated in Figure \ref{fig:wd_dist}. The two example planets in this figure correspond to some of the smallest ($1\,M_J$ at 2\,au) and largest ($13\,M_J$ at 3.91\,au) signals for which there is $S/N>3$ (almost $100\%$ detection probability). For $1\,M_J$ planets there are no $S/N>3$ candidates with magnitudes greater than 15.4 and beyond $30\,pc$. These high detection probability candidates are discussed further in Appendix \ref{app-table}. 

\begin{figure*}
    \centering
    \includegraphics[width=2\columnwidth]{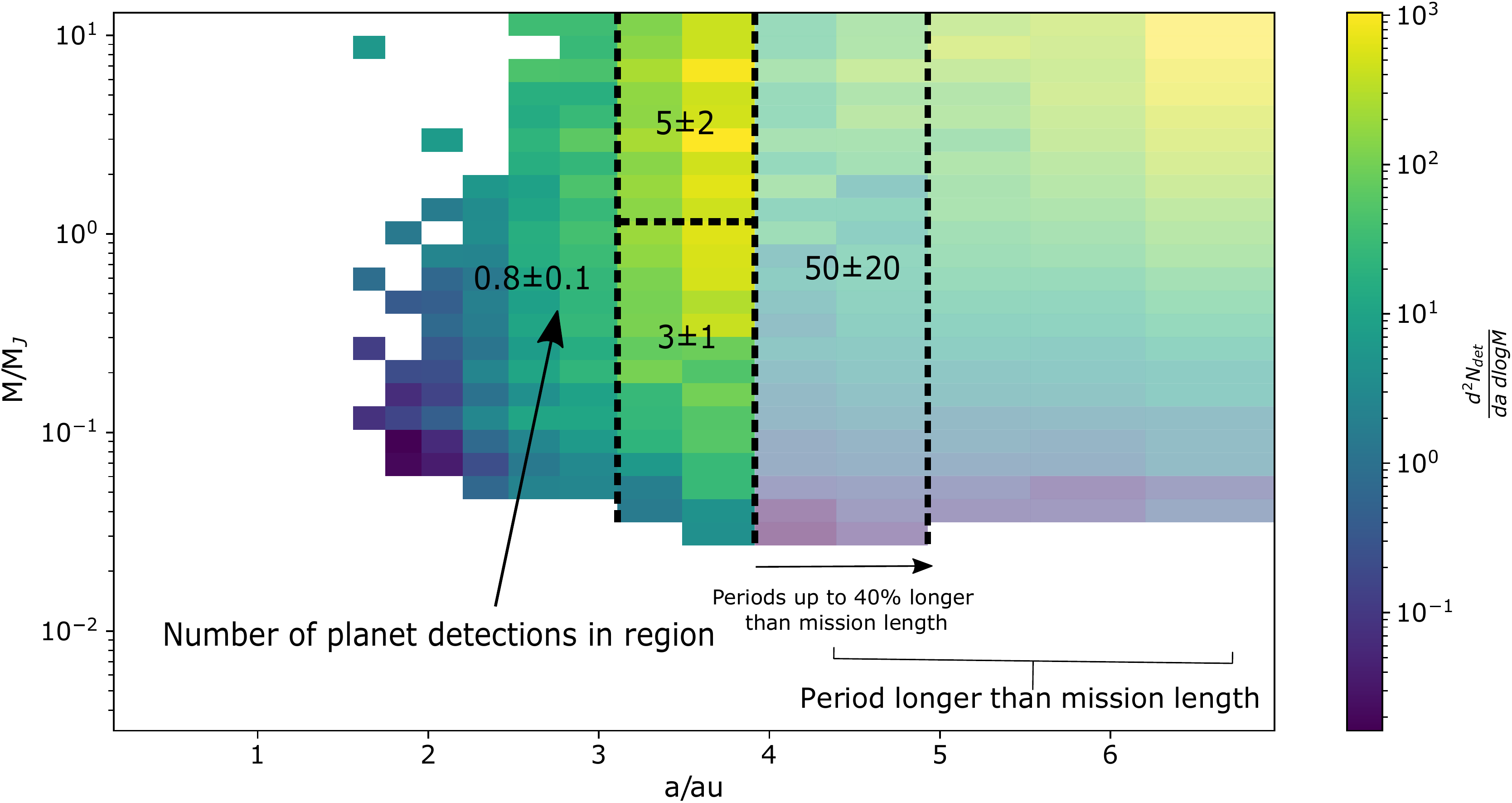}
    \caption{Predicted number of detected planets per semi-major axis and log M bin  ($\frac{\tm{d}^2N_{\tm{det}}}{\tm{d}a\:\tm{d\,log}M}$). Summing over the distribution gives the total number of planets we predict \Gaia will detect. The number of planets that we predict will be detected in each region is written in black. The vertical dashed boundary is at $a=3$\,au and the horizontal dashed boundary is at $M=M_J$. The faded region corresponds to planets with periods longer than the extended mission length, which have non-zero detection probabilities, but for which orbital parameters cannot be accurately recovered. The number of planet detections expected if periods up to 40\% longer than the mission length can be detected are shown in the bounded faded region (see \S \ref{dis-period}).} For periods longer than the extended mission length, detection probability also decreases so there is a fall off in $\frac{\tm{d}^2N_{\tm{det}}}{\tm{d}a\:\tm{d\,log}M}$. No planets less massive than $0.03\,M_J$ or interior to 1.6\,au will be detected. The gaps and isolated non-zero bins at small semi-major axes result from difference in tidal effects for planets at different semi-major axes motion of planets during post-main sequence evolution (see gaps in Figure \ref{fig:final_distribution1p0}).
    \label{fig:detected_planets}
\end{figure*}

We predict \Gaia will detect $8\pm2$ planets around white dwarfs that evolve due to mass loss and tides post-main sequence. Of these planets $9\pm2\,\%$  will lie inside 3 \,au (i.e. at most one), $90^{+10}_{-20}\,\%$ will lie between 3-3.91\,au. $40\pm10\,\%$ will be less massive than Jupiter. The predicted distribution of planets \Gaia can detect is shown in Figure \ref{fig:detected_planets}. We predict \Gaia will not be able detect planets less massive than $0.13\,M_J$ or with semi-major axes less than 1.6\,au. The faded region corresponds to planets with non-zero detection probabilities, but which have periods longer than the \Gaia extended mission length. The orbital parameters of these planets cannot be recovered reliably \citep{ranalli_astrometry_2018} so they are not included in the final predicted number of planet detections (see \S \ref{dis-period}) and the maximum semi-major axis where \Gaia can detect planets is 3.91\,au (for a $0.6\,M_\odot$ white dwarf).  The detected planets are concentrated at the largest masses and semi-major axes, because they have the strongest signals. Comparison with Figure \ref{fig:final_distribution1p0} highlights that \Gaia will not detect any of the low mass planets which should exist around white dwarfs, because they are below the \Gaia detection threshold. 

\section{Discussion}\label{discussion}

This work makes two key predictions: the number and distribution of planets \Gaia will find around white dwarfs that have evolved from the main-sequence by tides and mass loss. We predict \Gaia will find $8\pm2$ planets between 1.6-3.91\,au with masses between $0.03-13\,M_J$. Whilst the broad form of the distribution (mass and semi-major axis of discovered planets) is robust, the exact numbers of predicted planet detections depend crucially on the chosen main-sequence planet occurrence rate, the adopted post-main sequence planetary system evolution model and the ten year period cut-off for detected planets.

Nonetheless, we consider the numbers to be sufficiently reliable to be used as a benchmark comparison with \Gaia exoplanet detections in order to determine how important photoevaporation, common envelope evolution and dynamical scattering inwards are for planets around white dwarfs. 

\subsection{Validity of results}
\subsubsection{Post-main sequence planetary evolution model}\label{dis-model}
This work focuses only on the effects of stellar mass loss and tidal evolution on the fate of planets post-main sequence \citet{mustill_foretellings_2012}. Crucially, two processes are neglected which could potentially lead to planets on orbits interior to 1.6\,au, the innermost orbit predicted by tidal evolution in this work. Firstly, planets may survive inside and subsequently escape from the stellar envelope \citep[common envelope evolution][]{paczynski_common_1976}. While traditional common-envelope models predict a minimum mass of $10\mathrm{\,M_J}$ for the smallest planets that can survive common-envelope evolution \citep[e.g.,][]{nordhaus_2010}, the orbit of the Jupiter-mass planet WD 1856b can be explained by models that incorporate re-ionisation energy or successive engulfment events \citep{lagos_wd_2021,chamandy_successive_2021}. Common envelope evolution could drastically shrink the orbit, making the planet invisible to astrometry (but more detectable by the transit method). Secondly, multi-planet interactions or Kozai-Lidov interactions with stellar companions can lead to planets scattered inwards post-main sequence stellar evolution, as suggested by \citet{ronco_how_2020,oconnor_enhanced_2021,stephan_giant_2021}, amongst others. In the absence of tidal forces, scattering can decrease the semi-major axis by up to a factor of ~2 \citep[][Fig. 6]{mustill_long-term_2014}. The detection of planets interior to 1.6\,au (for which we already have the candidates around WD 1856 at 0.02\,au \citep{vanderburg_giant_2020} and WD 0141 at 0.18\,au \citep{arenou_gaia_2022}) provides an important test of the importance of common envelope evolution and/or dynamical scattering. 

The model also neglects photoevaporation, high stellar luminosity heating the planet's gaseous envelope such that it escapes, which could reduce a planet's size below the \textit{Gaia} observable limit \citep{villaver_can_2007}. \citet{villaver_can_2007} predict a strong dependence of photoevaporation on the planet's mass and orbital properties. $1M_J$ planets initially within 2\,au of their $1\,M_\odot$ progenitor host star lose 50$\%$ of their envelopes, whilst $5M_J$ planets initially at 3\,au only lost 0.4-5$\%$ of their envelope. Whilst the predicted \Gaia planet detections will suffer from small number statistics, if there is a lack of detections of $M_P\leq M_J$ planets it could be an indication that photoevaporation is important in these systems.

This work models tidal forces as the dissipation of an equilibrium tide by means of convective turbulence in the giant star's envelope \citep{mustill_foretellings_2012,zahn_tidal_1977}. The strength of this tide and scalings in the model (especially the dependence on tidal strength with the forcing frequency, i.e., the mean motion) are debated \citep[see, e.g.,][]{goldreich_nicholson_1977,zahn_tidal_1977,penev_2009,ogilvie_lesur_2012,ogilvie_tidal_2014}. \citet{mustill_foretellings_2012} found that the maximum orbital radius for engulfment can change by up to 1\,au for reasonable changes in these parameters. In principle, if we better understood the progenitor population of giant planets, decetection of planets around white dwarfs could be used to better constrain the tidal models. Unfortunately however, the number of planets \Gaia will detect around white dwarfs will be too small for this.

Another potential limitation to the models is the number of simulated planets. Increasing the number of simulated planets would reduce the fractional error on $R_{ijk}$, since $\sigma(R_{ijk})$ is Poissonian. This would be most significant for bins closest to the star, where the number of planets surviving in each bin is smallest. However, the predicted number of planet detections inside 3\,au is so low that increased resolution in this region would be an unnecessary computational expense. The number of planets per bin in the simulations sufficiently resolves the structure in the white dwarf planet distribution. 

\subsubsection{Planet occurrence rate}

The main-sequence planet occurrence rate contributes the most to the uncertainties in the predicted number of detected planets (see \S \ref{methods-error}). However, once \Gaia exoplanet detections are available for main-sequence stars, these will allow predictions for the number of planets \textit{Gaia} detect to be updated. This will further enable the use of the \Gaia white dwarf exoplanets to investigate the processes affecting planets' orbital evolution discussed in \S \ref{dis-model}.

Our chosen main-sequence planet occurrence rate from \citet{fernandes_hints_2019} crucially accounts for non-detections when calculating the planet occurrence rate by weighting the sum over the number of detected planets by the inverse of survey completeness at the mass and semi-major axis of each planet \citep[see Equation 1 in][]{fernandes_hints_2019}. However, estimates of completeness are still limited by the lack of planet observations beyond 1\,au where most \Gaia planet detections will lie. Short period planets are easily observed by the transit method and have been surveyed extensively by Kepler and TESS \citep{fulton_california-keplersurvey_2017,petigura_california-kepler_2017}. Planets beyond 1\,au have been discovered by radial velocity measurements e.g. \citet{mayor_harps_2011, wittenmyer_anglo-australian_2016}, microlensing e.g. \citet{gaudi_microlensing_2012} and direct imaging e.g. \citet{biller_gemininici_2013}, but these are are limited in number, relative to close-in planets.  

\subsubsection{Probing planets with periods longer than the \Gaia mission length}\label{dis-period}
In order to unambiguously detect and characterise the orbit of a planet, multiple orbital periods must be observed. This necessarily limits \Gaia to planets with periods shorter than the ten year mission length (interior to 3.91\,au). The faded region in Figure \ref{fig:detected_planets} predicts that a substantial population ($180^{+70}_{-80}$) of planets evolved from the main-sequence orbit white dwarfs into the four to eight au region. In this region the significant astrometric signals detected by \Gaias, would correspond to fractional orbital periods only. However, initial results from DR3 comparing spectroscopically and astrometrically determined periods \citep{dr3_documentation} suggests periods may be recoverable that are up to 20-40\% longer than the mission length. $50\pm20$ more planets could be found if detections between the mission length and this upper limit are possible (see the faded bounded region in Figure \ref{fig:detected_planets}). There are also other techniques which probe the population of planets on longer periods, based on astrometric detection of fractional orbital periods \citep[see e.g.][]{penoyre_binary_2020}.  
 
These planets have not been included in the predicted detected population in this work, but these planets would provide a population to compare in detail with individual post-main sequence planetary evolution models similar to comparisons of the Kepler radius valley with photoevaporation models e.g. \citep{owen_evaporation_2017}. These planets would be sufficient in number that it becomes possible to investigate how planet occurrence rate changes between more massive, early main-sequence stars (which have now reached the white dwarf phase) and lower mass main-sequence stars.

\subsubsection{Planets only survive around higher mass progenitors on orbital periods longer than the \Gaia mission length} \label{discussion-mass-cut}

Figure \ref{fig:final_distribution1p5} shows that planets end up beyond \Gaia detection limit in planetary systems with high mass progenitors. This is because the star loses a higher proportion of its mass and has stronger tidal forces. In this work, we attempt to account for the initial mass of the white dwarf progenitors, by using an initial-final mass relationship to determine which white dwarfs had high mass progenitors. No detectable planets are found around white dwarfs with initial masses higher than  $1.5\,M_\odot$, which using \citet{cummings_white_2018} corresponds to a final mass of $M=0.609\pm0.054M_\odot$ 

In practice, this meant that white dwarfs with masses higher than $0.663\,M_\odot$ were not included in the analysis presented in \S \ref{methods-number}. However, this mass cut off introduces uncertainty in the predicted number of detected planets, because there is error associated with the initial final mass relation and the mass values in the catalogue. 

The mixed masses from the \citet{gentilefusillo_catalogue_2021} catalogue were used in this work. These were derived by fitting white dwarf atmospheric models and evolutionary sequences to \Gaia photometry and photometric estimates of log g and $\boldmath{T_{\rm eff}}$ to obtain best fit values of mass and radius. Reddening for white dwarfs beyond 100\,pc was corrected for in the mass calculations. These estimates are unreliable for very high ($>1.3\,M_\odot$) or very low masses ($<0.2\,M_\odot$), but this is not a problem for the mass cut off as the very lowest masses are not underestimated enough to need excluding from the sample and the highest masses are far above the cut off. Using only \Gaia photometry introduces error in these masses, because as discussed in \citet{bergeron_measurement_2019} due to the high temperatures of white dwarfs it is important to include ‘bluer’ photometry such as SDSS ‘u’ to constrain the peak of the blackbody curve and the value of $\boldmath{T_{\rm eff}}$. White dwarfs in the catalogue that did not have masses were assumed to have masses of 0.6\,$M_\odot$ - the average mass of a white dwarf \citep{hollands_gaia_2018}. Excluding these white dwarfs instead does not change the predicted number of planet detections within error.

In order to predict which white dwarfs would have no planets on orbital periods shorter than the \Gaia mission length due to their high progenitor masses, it is necessary to deduce their evolutionary pathway and initial mass based on current observed properties, notably their mass (log g). For this process, we utilise initial to final mass relations which are notoriously unreliable. Uncertainties arise because two stellar models of different initial masses and metallicities or run through different stellar evolution codes can have the same final mass and it is difficult to obtain the metallicity of the progenitor mass from observations. 
In this work we used the \citet{cummings_white_2018} MIST-based initial-final mass relation based on stellar clusters, but the difference to the resulting predicted planet population if another relation \citep[e.g.][]{el-badry_empirical_2018,barrientos_improved_2021} were used is small.

The resolution of the post-main sequence stellar evolution models is also limited to $0.5\,M_\odot$ differences in progenitor mass. However, the slope of the initial final mass relation in the $1-1.5\,M_\odot$ progenitor range is so shallow and error on the relation so large that this outweighs the uncertainty due to model resolution.

The multiple sources of uncertainty on the mass cut off do not have a significant effect on the results, because the large number of white dwarfs mean the resulting fractional change in catalogue size is small. Repeating the analysis in this work using one and three $\sigma$ bounds on the white dwarf masses and the cut off value from the initial final mass relation to increase and decrease the number of white dwarfs in the catalogue do not change our results within the errors. Aside from the mathematics of the cut off, there is also uncertainty due to its initial validity. If we discover planets around white dwarfs with higher mass progenitors it suggests these planets have been scattered inwards or these white dwarfs have had a different evolutionary pathway. For example, if a white dwarf underwent a merger, the planetary systems may have a complex history, that proceeds in quite a different manner to the simple picture of tidal evolution presented in this work. 

\subsubsection{Other limitations}\label{dis-other}

In addition to the effects discussed above the \Gaia detection probabilities, the predicted magnitude of \Gaia DR5 errors and the width of the bins in mass-semi-major-axis space will have minor effects on the predicted number of planet detections. 
The \Gaia errors used in this work are predictions for DR5 based on EDR3 and we acknowledge this may be updated for future data releases. As the errors decrease with every data release this may lower the mass detection threshold. Revised detection probabilities may also change the exact number of detected planets.  

The assumption of circular orbits may also have a small effect on the predicted surviving planet distribution and the detected signal. Planets with initially eccentric orbits are more likely to be engulfed during post-main sequence evolution as their pericentres may pass inside the stellar envelope and eccentricity speeds up tidal semi-major axis decay. This could increase the final semi-major axis of the innermost surviving planet by around 0.25\,au for Jupiter mass planets \citep{mustill_foretellings_2012}. This may shift the inner boundary of the coloured region in Figure \ref{fig:detected_planets}, however so few planets are predicted to be detected interior to 3\,au that this will not make a significant difference to the total expected number of planets. Eccentricity also affects the detectability. Whilst planets that experienced tidal decay post-main sequence may have been circularised, multi-planet interactions may increase eccentricities in the white dwarf phase.

As already demonstrated in DR3, this astrometric detection method will also reveal larger companions to white dwarf, such as brown dwarfs or white dwarfs (double degenerates) \citep{arenou_gaia_2022}. This provides an opportunity to probe other white dwarf binary systems that have previously been difficult to detect. These are distinguishable from the planets discussed in this paper since the masses of these companions are larger than planets by an order of magnitude or more.

\subsection{Significance of the results} \label{discussion-significance}

This paper highlights the importance of \Gaia for the detection of exoplanets on medium orbital periods across the HR-diagram  \citep{perryman_astrometric_2014}.
White dwarf exoplanets are key to investigating the effects of stellar evolution on planetary systems. Whilst planetary systems around white dwarfs have been hypothesised for years to explain white dwarf pollution \citep{jura_extrasolar_2014}, there are few known planets  \citep{vanderburg_giant_2020,blackman_jovian_2021, luhman_discovery_2011}. Many unsuccessful surveys have searched for close-in planets orbiting white dwarfs \citep[e.g.][]{faedi_detection_2011,agol_transit_2011,burleigh_imaging_2002}, but no current techniques are sensitive to planets on wide (~year) orbits. This work highlights how \Gaia will shed light on these important planetary systems. 

Predictions are made for the planet population around white dwarfs, based on the evolution of the currently known main-sequence planet population due to tides and stellar mass loss, resulting in $8\pm2$ planets detectable by \Gaias. Notably, no planets are predicted interior to 1.6\,au, where they should have been engulfed by their host star, providing a test for the importance of common envelope evolution or dynamical scattering in the post-main sequence evolution of planetary systems.
If we see fewer Jupiter mass planets than expected it suggests they may be removed by photoevaporation. Existing observations and theoretical work  \citep[e.g.][]{veras_post-main-sequence_2016,vanderburg_giant_2020,munoz_kozai_2020,veras_detectable_2015,villaver_can_2007,alonso_transmission_2021,lagos_wd_2021} already suggest these processes could be important. If more planets are seen than expected it could provide evidence for second-generation planet formation. Predicting a distribution of planets unaffected by these processes can help us confirm which detected planets result from these processes and can be used to test models of these mechanisms. 

The results in this paper are broadly inline with previous work. The number of planet detections ($8\pm2$) predicted by this paper is slightly lower than the 13 predicted by \citet{perryman_astrometric_2014}. \citet{perryman_astrometric_2014} used the final mass of the white dwarf in a main-sequence planet occurrence rate, which may have overestimated the number of planets around white dwarfs in the \Gaia observable range, because it did not account for the expansion of planetary orbits and the engulfment of planets during post-main sequence stellar evolution.
The mass detection threshold, $0.03\pm0.004\,M_J$, is also lower than previous work ($2\,M_J$ \citet{silvotti_white_2011}). We attribute this to improved estimates of \Gaia errors and an enlarged catalogue of known white dwarfs. The lowest mass planets can only be found around the closest white dwarfs: including white dwarfs within 13\,pc lowers the mass detection threshold from $0.15\,M_J$ to $0.03\,M_J$. 

\subsection{Outer planets and the link with metals observed in the atmosphere of white dwarfs}  \label{dis-pollution}

Observations of polluted white dwarfs provide the best means of studying outer planetary systems to date. Most theories to explain the presence of metals in the atmospheres of white dwarfs suggest that material is scattered inwards from an outer planetary system that survived the star's evolution  \citep{farihi_circumstellar_2016}. The planets that \Gaia will detect provide a crucial piece of the puzzle: for the first time it will be possible to probe the outer planets orbiting white dwarfs. Here, we highlight that a link between the \Gaia planet detections and metals in the atmospheres of white dwarfs is not straight-forward and a one:one correlation is not anticipated. 

In order to investigate a potential link between \Gaia planet detections and asteroids or comets scattered inwards to pollute white dwarfs, we consider a simple scenario. Planets on the inner edges of planetesimal belts can scatter planetesimals and their ability to scatter planetesimals can increase following stellar mass loss. Many scattered bodies are ejected, but some can be scattered inwards, potentially ending up on star-grazing orbits close to the white dwarf \citep{bonsor_dynamical_2011, frewen_eccentric_2014,debes_link_2012}. Interior planets may be required to scatter planets along a chain \citep{marino_scattering_2018}. Many of the planets detected by \Gaia migrate less than expected from an adiabatic expansion due to stellar mass loss due to the effects of tides. If the planetary body directly exterior to them, be it a planet or a planetesimal belt, migrates outwards adiabatically, a dynamical `gap' in the planetary system is created, across which it becomes harder to scatter bodies inwards. Here we highlight that many planets detected by \Gaia will have created such dynamical `gaps' in their planetary systems, making the scattering of material inwards harder, and thus, the pollution of their host stars less likely. 

\begin{figure*}
    \centering
    \includegraphics[width=2\columnwidth]{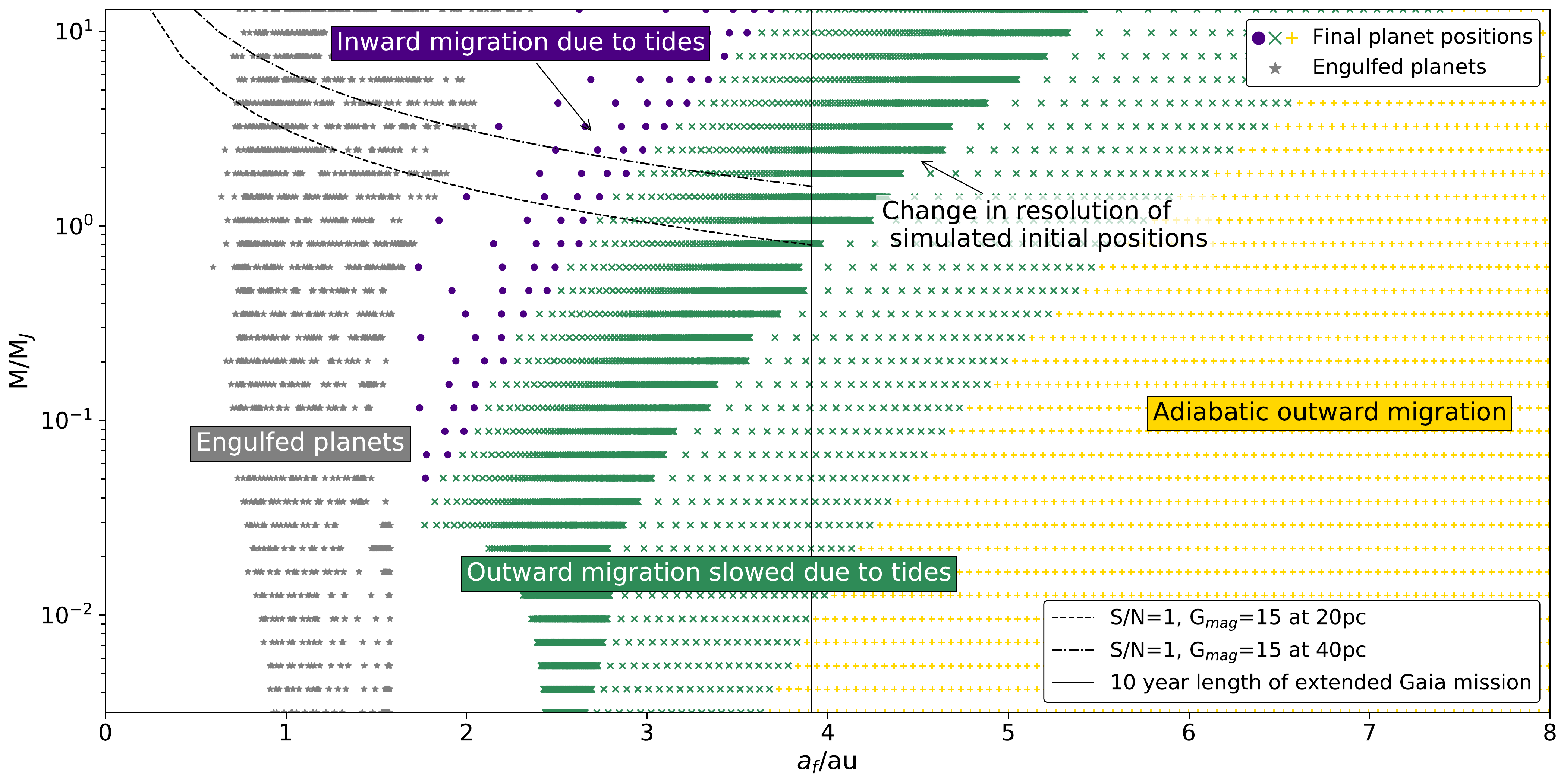}
    \caption{Final positions of simulated planets plotted as a function of their final semi-major axis and mass colour coded according to whether or not they could scatter material onto a white dwarf (see \S \ref{dis-pollution}). Engulfed planets are shown by the grey stars, inward migrating planets by the purple circles, planets moving outward adiabatically where material could be scattered from an outer planetesimal belt by the yellow pluses and outward moving planets with a dynamical `gap' in their system due to the effects of tides by the green crosses. The minimum planet mass detected as a function of semi-major axis for $S/N=1$ for a $G_{\tm{mag}}=15$ white dwarf at 20\,pc and 40\,pc  shown by the black dashed and dotted lines. The solid vertical line indicates the period cut off due to the length of the extended \textit{Gaia} mission. }
    \label{fig:summary_scatter}
\end{figure*}

To assess which planets in the predicted white dwarf planet distribution could contribute to white dwarf pollution we grouped the simulated planets into four categories (see Figure \ref{fig:summary_scatter}): planets engulfed by the star, planets that migrated outwards adiabatically and could scatter material from an outer planetesimal belt and planets whose orbits were influenced additionally by tides, either migrating inwards or outwards (less far than adiabatically). Whilst the inward migrating planets are less likely to be part of a planet chain scattering material onto the white dwarf from the outer planetary system, they could potentially interact with rocky material interior to their initial orbit, if the material migrates outwards adiabatically. For the mathematics of the categorisation see Appendix \ref{app-scat}. 

Almost all surviving planets in the \Gaia detection region have dynamical `gaps' in their planetary systems. This is shown in Figure \ref{fig:summary_scatter}, where only purple and green points lie above the \Gaia detection curves and inside 3.91\,au. All planets that migrate outwards adiabatically and could readily pollute white dwarfs by scattering material inwards lie beyond 3.91\,au and most of them are not massive enough to be detected. 
Even if \textit{Gaia} detects Jupiter mass planets that have migrated inwards, they are unlikely to be responsible for scattering planetary material from an outer system inwards. The detected Jupiters will have migrated inwards, separating themselves dynamically from the outer system. Additionally, Jupiter mass planets are more efficient at ejecting material from a planetary system  than scattering it inwards \citep{wyatt_how_2017}. They also scatter material too rapidly to give the observed long-term accretion of material onto some white dwarfs \citep{mustill_unstable_2018}. 
Low mass planets are more likely to be a long term source of pollution in a planetary system, because they scatter more material inwards and do so more slowly than Jupiter \citep{mustill_unstable_2018,oconnor_enhanced_2021}. These low mass planets are not detectable by \textit{Gaia}. Therefore, white dwarfs with \Gaia detected planets are less likely to be polluted than other white dwarfs with lower mass planets, that \Gaia does not detect. This is an important result: if \textit{Gaia} does not find a correlation between detected planets and polluted white dwarfs, this does not rule out the possibility that planets pollute white dwarfs. 

The white dwarf planet candidate in DR3 has been found around WD 0141-675, which is a polluted white dwarf \citep{debes_results_2010}. This planet is very close to its star (P=$33.65\pm0.05$\,d) so cannot have survived post-main sequence evolution as it is modelled in this work. It must have survived common envelope evolution or been scattered in during the white dwarf phase, as has been suggested for WD 1856b \citep{chamandy_successive_2021,lagos_wd_2021,merlov_red_2021,munoz_kozai_2020,oconnor_enhanced_2021}. A giant planet this close to a white dwarf would eject most material scattered inwards from an outer disc. An alternative explanation for the pollution might be the accretion of the planet's gaseous envelope, as suggested by \citet{gansicke_accretion_2019} for WDJ0914+1914. Planets on such short orbital periods are likely to be amongst the first detections, and it is not yet clear whether they are special cases or the norm.

\section{Conclusion}\label{conclusion}

\Gaia is revolutionary for almost every field in astrophysics, and exoplanets is no exception. Astrometric detection of wide-orbit (few au) planets orbiting stars across the HR diagram by \Gaia will uniquely probe the outer regions of planetary systems, crucially linked to the habitable zone around sun-like stars, such as our own. \Gaia planet detections around white dwarfs will probe the fate of planetary systems, providing secure evidence as to the key processes that influence planets post-main sequence. 

This work predicts that \Gaia will detect $8\pm2$ white dwarf exoplanets that evolved from the main-sequence only due to tides and the mass loss of the host star during post-main sequence evolution. These planets will have masses $0.03-13\,M_J$ and semi-major axes 1.6-3.91\,au. Comparison of these predictions with the \Gaia white dwarf planet detections will tell us how important dynamical scattering, common envelope evolution and photo-evaporation are. Notably, any planets detected interior to 1.6\,au, or around stars with progenitors more massive than $1.5\,M_\odot$ must have been subject to additional processes post-main sequence. 

Metals observed in the atmospheres of white dwarfs are commonly linked to outer planetary systems, in particular, comets or asteroids scattered inwards by planets. This work highlights how many white dwarfs planets detected by \Gaia will have evolved due to tides, leaving dynamical gaps in their outer planetary systems. These gaps render the scattering inwards of material to pollute the white dwarf harder. White dwarf pollution may, therefore, be more commonly associated with lower mass outer planets, hidden from \Gaia astrometry.

\section{Acknowledgements}
We would like to thank Piero Ranalli for sharing the exact values of his detection probabilities from \citet{ranalli_astrometry_2018}. We would also like to thank the reviewer for their insightful comments and  Laura Rogers, Zephyr Penoyre, Marc Brouwers, Andy Buchan, Elliot Lynch, Mark Wyatt and Rachel Fernandes and Giorgia Busso for their helpful discussions. AJM acknowledges support from the Swedish Research Council (grant 2017-04945). AB acknowledges funding from a Royal Society Dorothy Hodgkin Research Fellowship, DH150130 and a Royal Society University Research Fellowship, URF\textbackslash R1\textbackslash 211421. HS acknowledges funding on a NERC studentship NE/S007474/1 and an Exonian Graduate Scholarship from Exeter College, University of Oxford. For the purpose of open access, the authors have applied a Creative Commons Attribution (CC BY) licence to any Author Accepted Manuscript version arising.

\section{Data Availability}
Additional data and the code used in this publication are available on Github at \url{https://github.com/Hannah-RS/Gaia-white-dwarf}. The table of white dwarf candidates in Appendix \ref{app-table} will be available on Vizier after publication.

\bibliographystyle{mnras}
\bibliography{References} 

\newcommand{\SortNoop}[1]{}
\begin{thebibliography}{}
\makeatletter
\relax
\def\mn@urlcharsother{\let\do\@makeother \do\$\do\&\do\#\do\^\do\_\do\%\do\~}
\def\mn@doi{\begingroup\mn@urlcharsother \@ifnextchar [ {\mn@doi@}
  {\mn@doi@[]}}
\def\mn@doi@[#1]#2{\def\@tempa{#1}\ifx\@tempa\@empty \href
  {http://dx.doi.org/#2} {doi:#2}\else \href {http://dx.doi.org/#2} {#1}\fi
  \endgroup}
\def\mn@eprint#1#2{\mn@eprint@#1:#2::\@nil}
\def\mn@eprint@arXiv#1{\href {http://arxiv.org/abs/#1} {{\tt arXiv:#1}}}
\def\mn@eprint@dblp#1{\href {http://dblp.uni-trier.de/rec/bibtex/#1.xml}
  {dblp:#1}}
\def\mn@eprint@#1:#2:#3:#4\@nil{\def\@tempa {#1}\def\@tempb {#2}\def\@tempc
  {#3}\ifx \@tempc \@empty \let \@tempc \@tempb \let \@tempb \@tempa \fi \ifx
  \@tempb \@empty \def\@tempb {arXiv}\fi \@ifundefined
  {mn@eprint@\@tempb}{\@tempb:\@tempc}{\expandafter \expandafter \csname
  mn@eprint@\@tempb\endcsname \expandafter{\@tempc}}}

\bibitem[\protect\citeauthoryear{Agol}{Agol}{2011}]{agol_transit_2011}
Agol E.,  2011, \mn@doi [The Astrophysical Journal]
  {10.1088/2041-8205/731/2/L31}, 731, L31

\bibitem[\protect\citeauthoryear{Alonso et~al.,}{Alonso
  et~al.}{2021}]{alonso_transmission_2021}
Alonso R.,  et~al., 2021, \mn@doi [Astronomy \& Astrophysics]
  {10.1051/0004-6361/202140359}, 649, A131

\bibitem[\protect\citeauthoryear{Bailer-Jones, Rybizki, Fouesneau, Demleitner
  \& Andrae}{Bailer-Jones et~al.}{2021}]{bailer-jones_estimating_2021}
Bailer-Jones C. A.~L.,  Rybizki J.,  Fouesneau M.,  Demleitner M.,   Andrae R.,
   2021, \mn@doi [The Astronomical Journal] {10.3847/1538-3881/abd806}, 161,
  147

\bibitem[\protect\citeauthoryear{Barker}{Barker}{2020}]{barker_tidal_2020}
Barker A.~J.,  2020, \mn@doi [Monthly Notices of the Royal Astronomical
  Society] {10.1093/mnras/staa2405}, 498, 2270

\bibitem[\protect\citeauthoryear{Barnes \& Quinn}{Barnes \&
  Quinn}{2004}]{barnes_stability_2004}
Barnes R.,  Quinn T.,  2004, \mn@doi [The Astrophysical Journal]
  {10.1086/421321}, 611, 494

\bibitem[\protect\citeauthoryear{Barrientos \& Chanamé}{Barrientos \&
  Chanamé}{2021}]{barrientos_improved_2021}
Barrientos M.,  Chanamé J.,  2021, \mn@doi [The Astrophysical Journal]
  {10.3847/1538-4357/ac2f49}, 923, 181

\bibitem[\protect\citeauthoryear{Bear \& Soker}{Bear \&
  Soker}{2011}]{bear_evaporation_2011}
Bear E.,  Soker N.,  2011, \mn@doi [Monthly Notices of the Royal Astronomical
  Society] {10.1111/j.1365-2966.2011.18527.x}, 414, 1788

\bibitem[\protect\citeauthoryear{Belokurov et~al.,}{Belokurov
  et~al.}{2020}]{belokurov_unresolved_2020}
Belokurov V.,  et~al., 2020, \mn@doi [Monthly Notices of the Royal Astronomical
  Society] {10.1093/mnras/staa1522}, 496, 1922

\bibitem[\protect\citeauthoryear{Bergeron, Dufour, Fontaine, Coutu, Blouin,
  Genest-Beaulieu, Bédard  \& Rolland}{Bergeron
  et~al.}{2019}]{bergeron_measurement_2019}
Bergeron P.,  Dufour P.,  Fontaine G.,  Coutu S.,  Blouin S.,  Genest-Beaulieu
  C.,  Bédard A.,   Rolland B.,  2019, \mn@doi [The Astrophysical Journal]
  {10.3847/1538-4357/ab153a}, 876, 67

\bibitem[\protect\citeauthoryear{Bertolami}{Bertolami}{2018}]{bertolami_evolutionary_2018}
Bertolami M. M.~M.,  2018, \mn@doi [Proceedings of the International
  Astronomical Union] {10.1017/S1743921318007330}, 14, 36

\bibitem[\protect\citeauthoryear{Biller et~al.,}{Biller
  et~al.}{2013}]{biller_gemininici_2013}
Biller B.~A.,  et~al., 2013, \mn@doi [The Astrophysical Journal]
  {10.1088/0004-637X/777/2/160}, 777, 160

\bibitem[\protect\citeauthoryear{Blackman et~al.,}{Blackman
  et~al.}{2021}]{blackman_jovian_2021}
Blackman J.~W.,  et~al., 2021, \mn@doi [Nature] {10.1038/s41586-021-03869-6},
  598, 272

\bibitem[\protect\citeauthoryear{Bonsor, Mustill  \& Wyatt}{Bonsor
  et~al.}{2011}]{bonsor_dynamical_2011}
Bonsor A.,  Mustill A.~J.,   Wyatt M.~C.,  2011, \mn@doi [Monthly Notices of
  the Royal Astronomical Society] {10.1111/j.1365-2966.2011.18524.x}, 414, 930

\bibitem[\protect\citeauthoryear{Brown et~al.,}{Brown
  et~al.}{2021}]{brown_gaia_2021}
Brown A. G.~A.,  et~al., 2021, \mn@doi [Astronomy \& Astrophysics]
  {10.1051/0004-6361/202039657}, 649, A1

\bibitem[\protect\citeauthoryear{Bryan et~al.,}{Bryan
  et~al.}{2016}]{bryan_statistics_2016}
Bryan M.~L.,  et~al., 2016, \mn@doi [The Astrophysical Journal]
  {10.3847/0004-637X/821/2/89}, 821, 89

\bibitem[\protect\citeauthoryear{Burleigh, Clarke  \& Hodgkin}{Burleigh
  et~al.}{2002}]{burleigh_imaging_2002}
Burleigh M.~R.,  Clarke F.~J.,   Hodgkin S.~T.,  2002, \mn@doi [Monthly Notices
  of the Royal Astronomical Society] {10.1046/j.1365-8711.2002.05417.x}, 331,
  L41

\bibitem[\protect\citeauthoryear{Casertano et~al.,}{Casertano
  et~al.}{2008}]{casertano_double-blind_2008}
Casertano S.,  et~al., 2008, \mn@doi [Astronomy \& Astrophysics]
  {10.1051/0004-6361:20078997}, 482, 699

\bibitem[\protect\citeauthoryear{Chamandy, Blackman, Nordhaus  \&
  Wilson}{Chamandy et~al.}{2021}]{chamandy_successive_2021}
Chamandy L.,  Blackman E.~G.,  Nordhaus J.,   Wilson E.,  2021, \mn@doi
  [Monthly Notices of the Royal Astronomical Society: Letters]
  {10.1093/mnrasl/slab017}, 502, L110

\bibitem[\protect\citeauthoryear{Chiang, Kite, Kalas, Graham  \&
  Clampin}{Chiang et~al.}{2009}]{chiang_fomalhauts_2009}
Chiang E.,  Kite E.,  Kalas P.,  Graham J.~R.,   Clampin M.,  2009, \mn@doi
  [The Astrophysical Journal] {10.1088/0004-637X/693/1/734}, 693, 734

\bibitem[\protect\citeauthoryear{Cumming, Butler, Marcy, Vogt, Wright  \&
  Fischer}{Cumming et~al.}{2008}]{cumming_keck_2008}
Cumming A.,  Butler R.~P.,  Marcy G.~W.,  Vogt S.~S.,  Wright J.~T.,   Fischer
  D.~A.,  2008, \mn@doi [Publications of the Astronomical Society of the
  Pacific] {10.1086/588487}, 120, 531

\bibitem[\protect\citeauthoryear{Cummings, Kalirai, Tremblay, Ramirez-Ruiz  \&
  Choi}{Cummings et~al.}{2018}]{cummings_white_2018}
Cummings J.~D.,  Kalirai J.~S.,  Tremblay P.-E.,  Ramirez-Ruiz E.,   Choi J.,
  2018, \mn@doi [The Astrophysical Journal] {10.3847/1538-4357/aadfd6}, 866, 21

\bibitem[\protect\citeauthoryear{{Dawson}}{{Dawson}}{2018}]{Dawson}
{Dawson} R.~I.,  2018, in {Deeg} H.~J.,  {Belmonte} J.~A.,  eds, , Handbook of
  Exoplanets.
Springer International Publishing AG, p.~114,
  \mn@doi{10.1007/978-3-319-55333-7\_114}

\bibitem[\protect\citeauthoryear{Debes, Sigurdsson  \& Woodgate}{Debes
  et~al.}{2005}]{debes_cool_2005}
Debes J.~H.,  Sigurdsson S.,   Woodgate B.~E.,  2005, \mn@doi [The
  Astrophysical Journal] {10.1086/491640}, 633, 1168

\bibitem[\protect\citeauthoryear{Debes, Kilic, Werner  \& Rauch}{Debes
  et~al.}{2010}]{debes_results_2010}
Debes J.~H.,  Kilic M.,  Werner K.,   Rauch T.,  2010, in AIP Conference
  Proceedings, Volume 1273. Tubingen, (Germany), pp 488--491,
  \mn@doi{10.1063/1.3527870}, \url
  {http://aip.scitation.org/doi/abs/10.1063/1.3527870}

\bibitem[\protect\citeauthoryear{Debes, Walsh  \& Stark}{Debes
  et~al.}{2012}]{debes_link_2012}
Debes J.~H.,  Walsh K.~J.,   Stark C.,  2012, \mn@doi [The Astrophysical
  Journal] {10.1088/0004-637X/747/2/148}, 747, 148

\bibitem[\protect\citeauthoryear{Dufour, Blouin, Coutu, Fortin-Archambault,
  Thibeault, Bergeron  \& Fontaine}{Dufour et~al.}{2017}]{dufour_montreal_2017}
Dufour P.,  Blouin S.,  Coutu S.,  Fortin-Archambault M.,  Thibeault C.,
  Bergeron P.,   Fontaine G.,  2017, in 20th European White Dwarf Workshop.
  p.~3, \url {https://ui.adsabs.harvard.edu/abs/2017ASPC..509....3D}

\bibitem[\protect\citeauthoryear{El-Badry, Rix  \& Weisz}{El-Badry
  et~al.}{2018}]{el-badry_empirical_2018}
El-Badry K.,  Rix H.-W.,   Weisz D.~R.,  2018, \mn@doi [The Astrophysical
  Journal] {10.3847/2041-8213/aaca9c}, 860, L17

\bibitem[\protect\citeauthoryear{Faedi, West, Burleigh, Goad  \& Hebb}{Faedi
  et~al.}{2011}]{faedi_detection_2011}
Faedi F.,  West R.~G.,  Burleigh M.~R.,  Goad M.~R.,   Hebb L.,  2011, \mn@doi
  [Monthly Notices of the Royal Astronomical Society]
  {10.1111/j.1365-2966.2010.17488.x}, 410, 899

\bibitem[\protect\citeauthoryear{Farihi}{Farihi}{2016}]{farihi_circumstellar_2016}
Farihi J.,  2016, \mn@doi [New Astronomy Reviews]
  {10.1016/j.newar.2016.03.001}, 71, 9

\bibitem[\protect\citeauthoryear{Farihi, Wyatt, Greaves, Bonsor, Sibthorpe  \&
  Panić}{Farihi et~al.}{2014}]{farihi_alma_2014}
Farihi J.,  Wyatt M.~C.,  Greaves J.~S.,  Bonsor A.,  Sibthorpe B.,   Panić
  O.,  2014, \mn@doi [Monthly Notices of the Royal Astronomical Society]
  {10.1093/mnras/stu1545}, 444, 1821

\bibitem[\protect\citeauthoryear{Fernandes, Mulders, Pascucci, Mordasini  \&
  Emsenhuber}{Fernandes et~al.}{2019}]{fernandes_hints_2019}
Fernandes R.~B.,  Mulders G.~D.,  Pascucci I.,  Mordasini C.,   Emsenhuber A.,
  2019, \mn@doi [The Astrophysical Journal] {10.3847/1538-4357/ab0300}, 874, 81

\bibitem[\protect\citeauthoryear{Frewen \& Hansen}{Frewen \&
  Hansen}{2014}]{frewen_eccentric_2014}
Frewen S. F.~N.,  Hansen B. M.~S.,  2014, \mn@doi [Monthly Notices of the Royal
  Astronomical Society] {10.1093/mnras/stu097}, 439, 2442

\bibitem[\protect\citeauthoryear{Fulton et~al.,}{Fulton
  et~al.}{2014}]{fulton_search_2014}
Fulton B.~J.,  et~al., 2014, \mn@doi [The Astrophysical Journal]
  {10.1088/0004-637X/796/2/114}, 796, 114

\bibitem[\protect\citeauthoryear{Fulton et~al.,}{Fulton
  et~al.}{2017}]{fulton_california-keplersurvey_2017}
Fulton B.~J.,  et~al., 2017, \mn@doi [The Astronomical Journal]
  {10.3847/1538-3881/aa80eb}, 154, 109

\bibitem[\protect\citeauthoryear{Fulton et~al.,}{Fulton
  et~al.}{2021}]{fulton_california_2021}
Fulton B.~J.,  et~al., 2021, \mn@doi [The Astrophysical Journal Supplement
  Series] {10.3847/1538-4365/abfcc1}, 255, 14

\bibitem[\protect\citeauthoryear{{Gaia Collaboration} et~al.,}{{Gaia
  Collaboration} et~al.}{2016}]{gaia_collaboration_gaia_2016-1}
{Gaia Collaboration} et~al., 2016, \mn@doi [Astronomy and Astrophysics]
  {10.1051/0004-6361/201629512}, 595, A2

\bibitem[\protect\citeauthoryear{{Gaia Collaboration} et~al.,}{{Gaia
  Collaboration} et~al.}{2018}]{gaia_collaboration_gaia_2018}
{Gaia Collaboration} et~al., 2018, \mn@doi [Astronomy and Astrophysics]
  {10.1051/0004-6361/201833051}, 616, A1

\bibitem[\protect\citeauthoryear{{Gaia Collaboration} et~al.,}{{Gaia
  Collaboration} et~al.}{2021}]{klioner_gaia_2021}
{Gaia Collaboration} et~al., 2021, \mn@doi [Astronomy \& Astrophysics]
  {10.1051/0004-6361/202039734}, 649, A9

\bibitem[\protect\citeauthoryear{{Gaia Collaboration}, De~Ridder, Ripepi, Aerts
   \& {et al.}}{{Gaia Collaboration}
  et~al.}{2022a}]{asteroseismology_gaia_2022}
{Gaia Collaboration} De~Ridder J.,  Ripepi V.,  Aerts C.,   {et al.} 2022a,
  \mn@doi [Astronomy \& Astrophysics] {10.1051/0004-6361/202243767}

\bibitem[\protect\citeauthoryear{{Gaia Collaboration}, Arenou, Babusiaux  \&
  Barstow}{{Gaia Collaboration} et~al.}{2022b}]{arenou_gaia_2022}
{Gaia Collaboration} Arenou F.,  Babusiaux C.,   Barstow M.,  2022b, \mn@doi
  [Astronomy \& Astrophysics] {10.1051/0004-6361/202243782}

\bibitem[\protect\citeauthoryear{Gaudi}{Gaudi}{2012}]{gaudi_microlensing_2012}
Gaudi B.~S.,  2012, \mn@doi [Annual Review of Astronomy and Astrophysics]
  {10.1146/annurev-astro-081811-125518}, 50, 411

\bibitem[\protect\citeauthoryear{Gentile Fusillo et~al.,}{Gentile Fusillo
  et~al.}{2019}]{gentilefusillo_gaia_2019}
Gentile Fusillo N.~P.,  et~al., 2019, \mn@doi [Monthly Notices of the Royal
  Astronomical Society] {10.1093/mnras/sty3016}, 482, 4570

\bibitem[\protect\citeauthoryear{Gentile Fusillo et~al.,}{Gentile Fusillo
  et~al.}{2021}]{gentilefusillo_catalogue_2021}
Gentile Fusillo N.~P.,  et~al., 2021, \mn@doi [Monthly Notices of the Royal
  Astronomical Society] {10.1093/mnras/stab2672}, 508, 3877

\bibitem[\protect\citeauthoryear{{Goldreich} \& {Nicholson}}{{Goldreich} \&
  {Nicholson}}{1977}]{goldreich_nicholson_1977}
{Goldreich} P.,  {Nicholson} P.~D.,  1977, \mn@doi [\icarus]
  {10.1016/0019-1035(77)90163-4}, \href
  {https://ui.adsabs.harvard.edu/abs/1977Icar...30..301G} {30, 301}

\bibitem[\protect\citeauthoryear{Gould \& Kilic}{Gould \&
  Kilic}{2008}]{gould_finding_2008}
Gould A.,  Kilic M.,  2008, \mn@doi [The Astrophysical Journal]
  {10.1086/527476}, 673, L75

\bibitem[\protect\citeauthoryear{Gänsicke, Schreiber, Toloza, Gentile~Fusillo,
  Koester  \& Manser}{Gänsicke et~al.}{2019}]{gansicke_accretion_2019}
Gänsicke B.~T.,  Schreiber M.~R.,  Toloza O.,  Gentile~Fusillo N.~P.,  Koester
  D.,   Manser C.~J.,  2019, \mn@doi [Nature] {10.1038/s41586-019-1789-8}, 576,
  61

\bibitem[\protect\citeauthoryear{Harrison, Bonsor  \& Madhusudhan}{Harrison
  et~al.}{2018}]{harrison_polluted_2018}
Harrison J. H.~D.,  Bonsor A.,   Madhusudhan N.,  2018, \mn@doi [Monthly
  Notices of the Royal Astronomical Society] {10.1093/mnras/sty1700}, 479, 3814

\bibitem[\protect\citeauthoryear{{\SortNoop{Hippel}}~von Hippel \&
  {Thompson}}{{\SortNoop{Hippel}}~von Hippel \&
  {Thompson}}{2007}]{von_hippel_discovery_2007}
{\SortNoop{Hippel}}~von Hippel T.,  {Thompson} S.~E.,  2007, \mn@doi [The
  Astrophysical Journal] {10.1086/515434}, 661, 477

\bibitem[\protect\citeauthoryear{Hogan, Burleigh  \& Clarke}{Hogan
  et~al.}{2011}]{hogan_latest_2011}
Hogan E.,  Burleigh M.~R.,   Clarke F.~J.,  2011, in Planetary Systems Beyond
  The Main Sequence: Proceedings of the International Conference. AIP
  Conference Proceedings. pp 271--277, \mn@doi{10.1063/1.3556210}, \url
  {http://adsabs.harvard.edu/abs/2011AIPC.1331..271H}

\bibitem[\protect\citeauthoryear{Hollands, Tremblay, Gaensicke, Gentile-Fusillo
   \& Toonen}{Hollands et~al.}{2018}]{hollands_gaia_2018}
Hollands M.~A.,  Tremblay P.~E.,  Gaensicke B.~T.,  Gentile-Fusillo N.~P.,
  Toonen S.,  2018, \mn@doi [Monthly Notices of the Royal Astronomical Society]
  {10.1093/mnras/sty2057}, 480, 3942

\bibitem[\protect\citeauthoryear{Ivanova et~al.,}{Ivanova
  et~al.}{2013}]{ivanova_common_2013}
Ivanova N.,  et~al., 2013, \mn@doi [The Astronomy and Astrophysics Review]
  {10.1007/s00159-013-0059-2}, 21, 59

\bibitem[\protect\citeauthoryear{Jura}{Jura}{2003}]{jura_tidally_2003}
Jura M.,  2003, \mn@doi [The Astrophysical Journal] {10.1086/374036}, 584, L91

\bibitem[\protect\citeauthoryear{Jura \& Young}{Jura \&
  Young}{2014}]{jura_extrasolar_2014}
Jura M.,  Young E.,  2014, \mn@doi [Annual Review of Earth and Planetary
  Sciences] {10.1146/annurev-earth-060313-054740}, 42, 45

\bibitem[\protect\citeauthoryear{Kaltenegger, MacDonald, Kozakis, Lewis,
  Mamajek, McDowell  \& Vanderburg}{Kaltenegger
  et~al.}{2020}]{kaltenegger_white_2020}
Kaltenegger L.,  MacDonald R.~J.,  Kozakis T.,  Lewis N.~K.,  Mamajek E.~E.,
  McDowell J.~C.,   Vanderburg A.,  2020, \mn@doi [The Astrophysical Journal]
  {10.3847/2041-8213/aba9d3}, 901, L1

\bibitem[\protect\citeauthoryear{Kervella, Arenou, Mignard  \&
  Thévenin}{Kervella et~al.}{2019}]{kervella_stellar_2019}
Kervella P.,  Arenou F.,  Mignard F.,   Thévenin F.,  2019, \mn@doi [Astronomy
  and Astrophysics] {10.1051/0004-6361/201834371}, 623, A72

\bibitem[\protect\citeauthoryear{Kervella, Arenou  \& Thévenin}{Kervella
  et~al.}{2022}]{kervella_stellar_2022}
Kervella P.,  Arenou F.,   Thévenin F.,  2022, \mn@doi [Astronomy \&
  Astrophysics] {10.1051/0004-6361/202142146}, 657, A7

\bibitem[\protect\citeauthoryear{Kuchner, Koresko  \& Brown}{Kuchner
  et~al.}{1998}]{kuchner_keck_1998}
Kuchner M.~J.,  Koresko C.~D.,   Brown M.~E.,  1998, \mn@doi [The Astrophysical
  Journal] {10.1086/311725}, 508, L81

\bibitem[\protect\citeauthoryear{Lagos, Schreiber, Zorotovic, Gänsicke, Ronco
  \& Hamers}{Lagos et~al.}{2021}]{lagos_wd_2021}
Lagos F.,  Schreiber M.~R.,  Zorotovic M.,  Gänsicke B.~T.,  Ronco M.~P.,
  Hamers A.~S.,  2021, \mn@doi [Monthly Notices of the Royal Astronomical
  Society] {10.1093/mnras/staa3703}, 501, 676

\bibitem[\protect\citeauthoryear{Lindegren}{Lindegren}{2018}]{LL:LL-124}
Lindegren L.,  2018, {R}e-normalising the astrometric chi-square in {G}aia
  {D}{R}2, GAIA-C3-TN-LU-LL-124, \url
  {https://www.cosmos.esa.int/web/gaia/public-dpac-documents}

\bibitem[\protect\citeauthoryear{Loeb \& Maoz}{Loeb \&
  Maoz}{2013}]{loeb_detecting_2013}
Loeb A.,  Maoz D.,  2013, \mn@doi [Monthly Notices of the Royal Astronomical
  Society] {10.1093/mnrasl/slt026}, 432, L11

\bibitem[\protect\citeauthoryear{Luhman, Burgasser  \& Bochanski}{Luhman
  et~al.}{2011}]{luhman_discovery_2011}
Luhman K.~L.,  Burgasser A.~J.,   Bochanski J.~J.,  2011, \mn@doi [The
  Astrophysical Journal Letters] {10.1088/2041-8205/730/1/L9}, 730, L9

\bibitem[\protect\citeauthoryear{Marino, Bonsor, Wyatt  \& Kral}{Marino
  et~al.}{2018}]{marino_scattering_2018}
Marino S.,  Bonsor A.,  Wyatt M.~C.,   Kral Q.,  2018, \mn@doi [Monthly Notices
  of the Royal Astronomical Society] {10.1093/mnras/sty1475}, 479, 1651

\bibitem[\protect\citeauthoryear{Matthews \& Claussen}{Matthews \&
  Claussen}{2018}]{matthews_evolved_2018}
Matthews L.~D.,  Claussen M.~J.,  2018, in Science with a Next Generation Very
  Large Array, ASP Conference Series, Vol. 517. ASP Monograph 7. Edited by Eric
  Murphy. p.~281, \url
  {https://ui.adsabs.harvard.edu/abs/2018ASPC..517..281M/abstract}

\bibitem[\protect\citeauthoryear{Mayor et~al.,}{Mayor
  et~al.}{2011}]{mayor_harps_2011}
Mayor M.,  et~al., 2011, arXiv e-prints, 1109, arXiv:1109.2497

\bibitem[\protect\citeauthoryear{Merlov, Bear  \& Soker}{Merlov
  et~al.}{2021}]{merlov_red_2021}
Merlov A.,  Bear E.,   Soker N.,  2021, \mn@doi [The Astrophysical Journal]
  {10.3847/2041-8213/ac0f7d}, 915, L34

\bibitem[\protect\citeauthoryear{{Morbidelli}, {Levison}  \&
  {Gomes}}{{Morbidelli} et~al.}{2008}]{Morbidelli2008}
{Morbidelli} A.,  {Levison} H.~F.,   {Gomes} R.,  2008, in {Barucci} M.~A.,
  {Boehnhardt} H.,  {Cruikshank} D.~P.,  {Morbidelli} A.,   {Dotson} R.,  eds,
  , The Solar System Beyond Neptune.
American Astronomical Society, DDA meeting \#39, id.12.04, p.~275, \url
  {https://ui.adsabs.harvard.edu/abs/2008ssbn.book..275M}

\bibitem[\protect\citeauthoryear{Mustill \& Villaver}{Mustill \&
  Villaver}{2012}]{mustill_foretellings_2012}
Mustill A.~J.,  Villaver E.,  2012, \mn@doi [The Astrophysical Journal]
  {10.1088/0004-637X/761/2/121}, 761, 121

\bibitem[\protect\citeauthoryear{Mustill, Veras  \& Villaver}{Mustill
  et~al.}{2014}]{mustill_long-term_2014}
Mustill A.~J.,  Veras D.,   Villaver E.,  2014, \mn@doi [Monthly Notices of the
  Royal Astronomical Society] {10.1093/mnras/stt1973}, 437, 1404

\bibitem[\protect\citeauthoryear{Mustill, Villaver, Veras, Gänsicke  \&
  Bonsor}{Mustill et~al.}{2018}]{mustill_unstable_2018}
Mustill A.~J.,  Villaver E.,  Veras D.,  Gänsicke B.~T.,   Bonsor A.,  2018,
  \mn@doi [Monthly Notices of the Royal Astronomical Society]
  {10.1093/mnras/sty446}, 476, 3939

\bibitem[\protect\citeauthoryear{Muñoz \& Petrovich}{Muñoz \&
  Petrovich}{2020}]{munoz_kozai_2020}
Muñoz D.~J.,  Petrovich C.,  2020, \mn@doi [The Astrophysical Journal Letters]
  {10.3847/2041-8213/abc564}, 904, L3

\bibitem[\protect\citeauthoryear{{Nordhaus}, {Spiegel}, {Ibgui}, {Goodman}  \&
  {Burrows}}{{Nordhaus} et~al.}{2010}]{nordhaus_2010}
{Nordhaus} J.,  {Spiegel} D.~S.,  {Ibgui} L.,  {Goodman} J.,   {Burrows} A.,
  2010, \mn@doi [Monthly Notices of the Royal Astronomical Society]
  {10.1111/j.1365-2966.2010.17155.x}, \href
  {https://ui.adsabs.harvard.edu/abs/2010MNRAS.408..631N} {408, 631}

\bibitem[\protect\citeauthoryear{Ogilvie}{Ogilvie}{2014}]{ogilvie_tidal_2014}
Ogilvie G.~I.,  2014, \mn@doi [Annual Review of Astronomy and Astrophysics]
  {10.1146/annurev-astro-081913-035941}, 52, 171

\bibitem[\protect\citeauthoryear{{Ogilvie} \& {Lesur}}{{Ogilvie} \&
  {Lesur}}{2012}]{ogilvie_lesur_2012}
{Ogilvie} G.~I.,  {Lesur} G.,  2012, \mn@doi [Monthly Notices of the Royal
  Astronomical Society] {10.1111/j.1365-2966.2012.20630.x}, \href
  {https://ui.adsabs.harvard.edu/abs/2012MNRAS.422.1975O} {422, 1975}

\bibitem[\protect\citeauthoryear{Owen \& Wu}{Owen \&
  Wu}{2017}]{owen_evaporation_2017}
Owen J.~E.,  Wu Y.,  2017, \mn@doi [The Astrophysical Journal]
  {10.3847/1538-4357/aa890a}, 847, 29

\bibitem[\protect\citeauthoryear{O’Connor, Liu  \& Lai}{O’Connor
  et~al.}{2021}]{oconnor_enhanced_2021}
O’Connor C.~E.,  Liu B.,   Lai D.,  2021, \mn@doi [Monthly Notices of the
  Royal Astronomical Society] {10.1093/mnras/staa3723}, 501, 507

\bibitem[\protect\citeauthoryear{Paczynski}{Paczynski}{1976}]{paczynski_common_1976}
Paczynski B.,  1976, \mn@doi [Symposium - International Astronomical Union]
  {10.1017/S0074180900011864}, 73, 75

\bibitem[\protect\citeauthoryear{{Penev}, {Sasselov}, {Robinson}  \&
  {Demarque}}{{Penev} et~al.}{2009}]{penev_2009}
{Penev} K.,  {Sasselov} D.,  {Robinson} F.,   {Demarque} P.,  2009, \mn@doi
  [\apj] {10.1088/0004-637X/704/2/930}, \href
  {https://ui.adsabs.harvard.edu/abs/2009ApJ...704..930P} {704, 930}

\bibitem[\protect\citeauthoryear{Penoyre, Belokurov, Wyn Evans, Everall  \&
  Koposov}{Penoyre et~al.}{2020}]{penoyre_binary_2020}
Penoyre Z.,  Belokurov V.,  Wyn Evans N.,  Everall A.,   Koposov S.~E.,  2020,
  \mn@doi [Monthly Notices of the Royal Astronomical Society]
  {10.1093/mnras/staa1148}, 495, 321

\bibitem[\protect\citeauthoryear{Penoyre, Belokurov  \& Evans}{Penoyre
  et~al.}{2022}]{penoyre_astrometric_2022}
Penoyre Z.,  Belokurov V.,   Evans N.~W.,  2022, \mn@doi [Monthly Notices of
  the Royal Astronomical Society] {10.1093/mnras/stac959}, pp 2437--2456

\bibitem[\protect\citeauthoryear{Perryman, Hartman, Bakos  \&
  Lindegren}{Perryman et~al.}{2014}]{perryman_astrometric_2014}
Perryman M.,  Hartman J.,  Bakos G.~a.,   Lindegren L.,  2014, \mn@doi [The
  Astrophysical Journal] {10.1088/0004-637X/797/1/14}, 797, 14

\bibitem[\protect\citeauthoryear{Petigura et~al.,}{Petigura
  et~al.}{2017}]{petigura_california-kepler_2017}
Petigura E.~A.,  et~al., 2017, \mn@doi [The Astronomical Journal]
  {10.3847/1538-3881/aa80de}, 154, 107

\bibitem[\protect\citeauthoryear{Pourbaix et~al.,}{Pourbaix
  et~al.}{2022}]{dr3_documentation}
Pourbaix D.,  et~al., 2022, Gaia DR3 Documentation: Chapter 7 Non-single stars;
  Combined solutions and post-treatment - Quality assessment and validation,
  \url
  {https://gea.esac.esa.int/archive/documentation/GDR3/Data_analysis/chap_cu4nss/sec_cu4nss_combined/ssec_cu4nss_combined_quality.html}

\bibitem[\protect\citeauthoryear{Ranalli, Hobbs  \& Lindegren}{Ranalli
  et~al.}{2018}]{ranalli_astrometry_2018}
Ranalli P.,  Hobbs D.,   Lindegren L.,  2018, \mn@doi [Astronomy \&
  Astrophysics] {10.1051/0004-6361/201730921}, 614, A30

\bibitem[\protect\citeauthoryear{Ronco, Schreiber, Giuppone, Veras, Cuadra  \&
  Guilera}{Ronco et~al.}{2020}]{ronco_how_2020}
Ronco M.~P.,  Schreiber M.~R.,  Giuppone C.~A.,  Veras D.,  Cuadra J.,
  Guilera O.~M.,  2020, \mn@doi [The Astrophysical Journal Letters]
  {10.3847/2041-8213/aba35f}, 898, L23

\bibitem[\protect\citeauthoryear{Santerne et~al.,}{Santerne
  et~al.}{2016}]{santerne_sophie_2016}
Santerne A.,  et~al., 2016, \mn@doi [Astronomy and Astrophysics]
  {10.1051/0004-6361/201527329}, 587, A64

\bibitem[\protect\citeauthoryear{{Shannon}, {Bonsor}, {Kral}  \&
  {Matthews}}{{Shannon} et~al.}{2016}]{Shannon2016}
{Shannon} A.,  {Bonsor} A.,  {Kral} Q.,   {Matthews} E.,  2016, \mn@doi
  [Monthly Notices of the Royal Astronomical Society] {10.1093/mnrasl/slw143},
  \href {https://ui.adsabs.harvard.edu/abs/2016MNRAS.462L.116S} {462, L116}

\bibitem[\protect\citeauthoryear{Sigurdsson, Richer, Hansen, Stairs  \&
  Thorsett}{Sigurdsson et~al.}{2003}]{sigurdsson_young_2003}
Sigurdsson S.,  Richer H.~B.,  Hansen B.~M.,  Stairs I.~H.,   Thorsett S.~E.,
  2003, \mn@doi [Science] {10.1126/science.1086326}, 301, 193

\bibitem[\protect\citeauthoryear{Silvotti, Sozzetti  \& Lattanzi}{Silvotti
  et~al.}{2011}]{silvotti_white_2011}
Silvotti R.,  Sozzetti A.,   Lattanzi M.,  2011, \mn@doi [AIP Conference
  Proceedings] {10.1063/1.3556222}, 1331, 336

\bibitem[\protect\citeauthoryear{Stephan, Naoz  \& Gaudi}{Stephan
  et~al.}{2021}]{stephan_giant_2021}
Stephan A.~P.,  Naoz S.,   Gaudi B.~S.,  2021, \mn@doi [The Astrophysical
  Journal] {10.3847/1538-4357/ac22a9}, 922, 4

\bibitem[\protect\citeauthoryear{Swan, Kenyon, Farihi, Dennihy, Gänsicke,
  Hermes, Melis  \& von Hippel}{Swan et~al.}{2021}]{swan_collisions_2021}
Swan A.,  Kenyon S.~J.,  Farihi J.,  Dennihy E.,  Gänsicke B.~T.,  Hermes
  J.~J.,  Melis C.,   von Hippel T.,  2021, \mn@doi [Monthly Notices of the
  Royal Astronomical Society] {10.1093/mnras/stab1738}, 506, 432

\bibitem[\protect\citeauthoryear{Tokunaga, Becklin  \& Zuckerman}{Tokunaga
  et~al.}{1990}]{tokunaga_infrared_1990}
Tokunaga A.~T.,  Becklin E.~E.,   Zuckerman B.,  1990, \mn@doi [The
  Astrophysical Journal] {10.1086/185770}, 358, L21

\bibitem[\protect\citeauthoryear{Vanderburg et~al.,}{Vanderburg
  et~al.}{2020}]{vanderburg_giant_2020}
Vanderburg A.,  et~al., 2020, \mn@doi [Nature] {10.1038/s41586-020-2713-y},
  585, 363

\bibitem[\protect\citeauthoryear{{Vassiliadis} \& {Wood}}{{Vassiliadis} \&
  {Wood}}{1993}]{vassiliadis_wood_1993}
{Vassiliadis} E.,  {Wood} P.~R.,  1993, \mn@doi [\apj] {10.1086/173033}, \href
  {https://ui.adsabs.harvard.edu/abs/1993ApJ...413..641V} {413, 641}

\bibitem[\protect\citeauthoryear{Veras}{Veras}{2016}]{veras_post-main-sequence_2016}
Veras D.,  2016, \mn@doi [Royal Society Open Science] {10.1098/rsos.150571}, 3,
  150571

\bibitem[\protect\citeauthoryear{Veras \& Gänsicke}{Veras \&
  Gänsicke}{2015}]{veras_detectable_2015}
Veras D.,  Gänsicke B.~T.,  2015, \mn@doi [Monthly Notices of the Royal
  Astronomical Society] {10.1093/mnras/stu2475}, 447, 1049

\bibitem[\protect\citeauthoryear{Veras, Wyatt, Mustill, Bonsor  \&
  Eldridge}{Veras et~al.}{2011}]{veras_great_2011}
Veras D.,  Wyatt M.~C.,  Mustill A.~J.,  Bonsor A.,   Eldridge J.~J.,  2011,
  \mn@doi [Monthly Notices of the Royal Astronomical Society]
  {10.1111/j.1365-2966.2011.19393.x}, 417, 2104

\bibitem[\protect\citeauthoryear{Veras, Mustill, Bonsor  \& Wyatt}{Veras
  et~al.}{2013}]{veras_simulations_2013}
Veras D.,  Mustill A.~J.,  Bonsor A.,   Wyatt M.~C.,  2013, \mn@doi [Monthly
  Notices of the Royal Astronomical Society] {10.1093/mnras/stt289}, 431, 1686

\bibitem[\protect\citeauthoryear{Villaver \& Livio}{Villaver \&
  Livio}{2007}]{villaver_can_2007}
Villaver E.,  Livio M.,  2007, \mn@doi [The Astrophysical Journal]
  {10.1086/516746}, 661, 1192

\bibitem[\protect\citeauthoryear{Wisdom}{Wisdom}{1980}]{wisdom_jack_resonance_1980}
Wisdom J.,  1980, The Astronomical Journal, 85, 1122

\bibitem[\protect\citeauthoryear{Wittenmyer et~al.,}{Wittenmyer
  et~al.}{2016}]{wittenmyer_anglo-australian_2016}
Wittenmyer R.~A.,  et~al., 2016, \mn@doi [The Astrophysical Journal]
  {10.3847/0004-637X/819/1/28}, 819, 28

\bibitem[\protect\citeauthoryear{Wyatt, Bonsor, Jackson, Marino  \&
  Shannon}{Wyatt et~al.}{2017}]{wyatt_how_2017}
Wyatt M.~C.,  Bonsor A.,  Jackson A.~P.,  Marino S.,   Shannon A.,  2017,
  \mn@doi [Monthly Notices of the Royal Astronomical Society]
  {10.1093/mnras/stw2633}, 464, 3385

\bibitem[\protect\citeauthoryear{Xu, Ertel, Wahhaj, Milli, Scicluna  \&
  Bertrang}{Xu et~al.}{2015}]{xu_extreme-ao_2015}
Xu S.,  Ertel S.,  Wahhaj Z.,  Milli J.,  Scicluna P.,   Bertrang G. H.-M.,
  2015, \mn@doi [Astronomy \& Astrophysics] {10.1051/0004-6361/201526179}, 579,
  L8

\bibitem[\protect\citeauthoryear{Zahn}{Zahn}{1977}]{zahn_tidal_1977}
Zahn J.-P.,  1977, Astronomy and Astrophysics, 57, 383

\bibitem[\protect\citeauthoryear{Zhu \& Dong}{Zhu \&
  Dong}{2021}]{zhu_exoplanet_2021}
Zhu W.,  Dong S.,  2021, \mn@doi [Annual Review of Astronomy and Astrophysics]
  {10.1146/annurev-astro-112420-020055}, 59, 291

\bibitem[\protect\citeauthoryear{Zuckerman, Koester, Dufour, Melis, Klein  \&
  Jura}{Zuckerman et~al.}{2011}]{zuckerman_aluminumcalcium-rich_2011}
Zuckerman B.,  Koester D.,  Dufour P.,  Melis C.,  Klein B.,   Jura M.,  2011,
  \mn@doi [The Astrophysical Journal] {10.1088/0004-637X/739/2/101}, 739, 101

\makeatother
\end{thebibliography}


\appendix
\section{Effects of tides}\label{app-tides}

\begin{figure*}
    \centering
    \includegraphics[width=2\columnwidth]{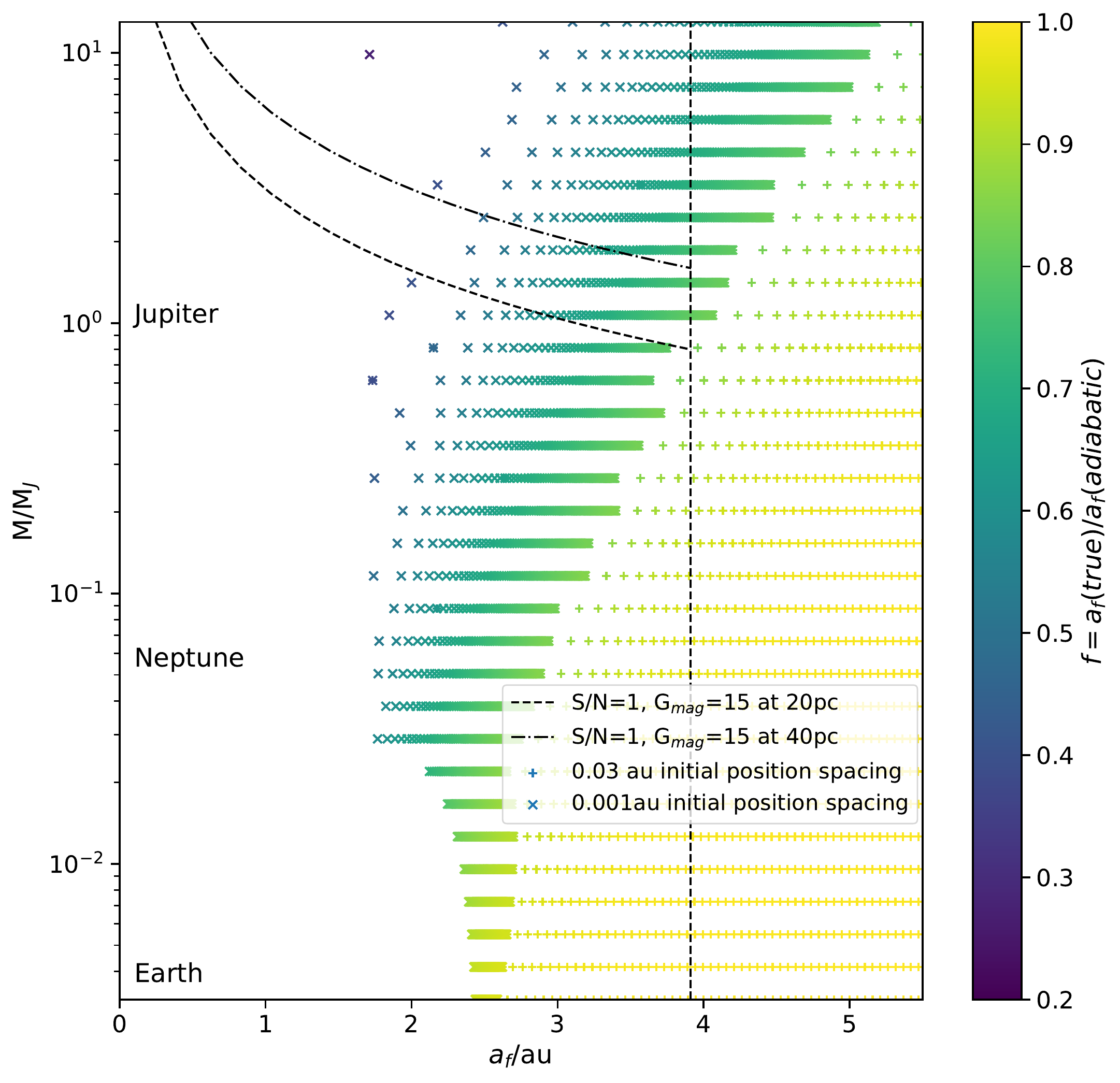}
    \caption{Final positions of simulated planets with periods less than ten years colour coded by $f = \frac{a_\tm{f}(\tm{true})}{a_\tm{f}(\tm{adiabatic})}$. Yellow points indicate planets' orbits which underwent an adiabatic expansion, whilst green to blue points indicate planets affected by tidal forces. Values of $f<0.57$ indicate the planet moved inwards.}
    \label{fig:effects_of_tides}
\end{figure*}

\begin{figure*}
    \centering
    \includegraphics[width=2\columnwidth]{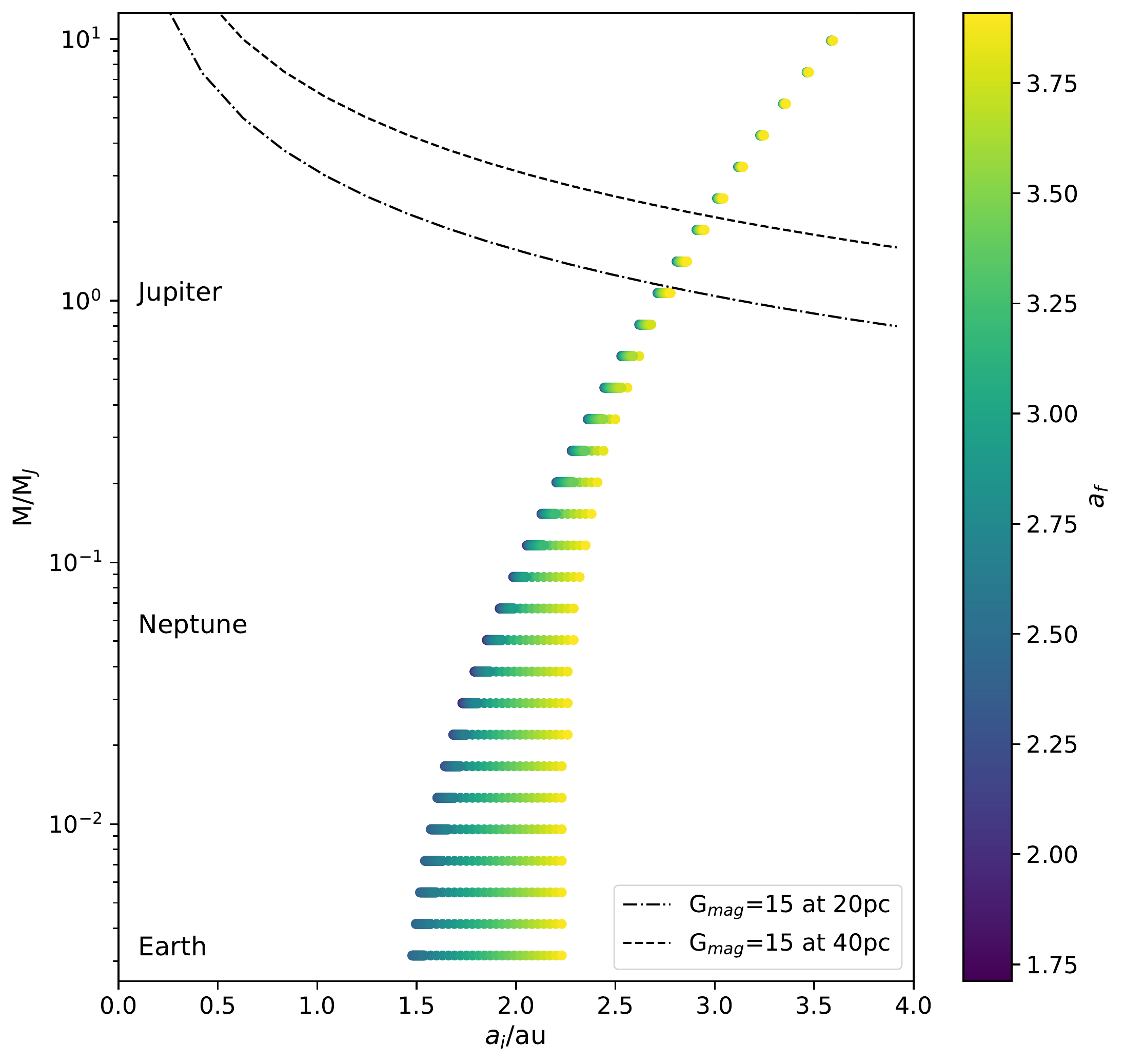}
    \caption{Initial positions of planets with final positions inside the \Gaia detection region colour-coded by final position.}
    \label{fig:pos_in_range}
\end{figure*}

As discussed in \S \ref{results-planet-distribution} and \S \ref{dis-model} tides have a strong effect on the final position of planets above a certain mass. The strength of this effect can be quantified by $f$, the ratio of the final position of a planet and the final position expected from an adiabatic expansion: \begin{equation}
    f = \frac{a_\tm{f}(\tm{true})}{a_\tm{f}(\tm{adiabatic})}.
\end{equation}
The final position of a planet can be written as \begin{equation}
    a_\tm{f}=\frac{M_\star(\tm{initial})}{M_\star(\tm{final})}fa_\tm{i} =pfa_\tm{i}.
\end{equation} If $f<\frac{1}{p}$ then $a_\tm{f}<a_\tm{i}$ and tidal forces were strong enough that the planet moved inwards. For a $1M-\odot$ star in our simulations $p=0.5702$. Planets closer to their star experience stronger tidal forces and more massive planets experience tidal forces over a wider range of initial positions, as shown by Figure \ref{fig:effects_of_tides}. These tidal forces cause planets with a small range of initial positions (Figure \ref{fig:pos_in_range}) to spread over a wide range of final positions. This contributes to the low number of predicted planet detections, because the semi-major axis range of main-sequence planets which will evolve to have periods less than ten years (the predicted \Gaia extended mission length) is small.


\section{Dynamical Gap Calculations}\label{app-scat}
\textit{For definitions of $f$ and $p$ see Appendix \ref{app-tides}}.

Planetary systems tend on average to be dynamically packed, or in other words the planetary bodies settle down to a stable configuration in which they are sufficiently separated that interactions occur on timescales longer than the lifetime of the system \citep{barnes_stability_2004,Shannon2016,Dawson}. The inner edge of a planetesimal belt can be carved by a planet that is orbiting just interior to it \citep{chiang_fomalhauts_2009}, just like Neptune sculpts the inner edge of the Kuiper belt \citep[\eg][]{Morbidelli2008}. If this is the case, when the star undergoes mass loss, the planetesimals in the belt migrate outwards adiabatically, but the planet interior to the belt may also be influenced by tides from the AGB star. As a result, the planet migrates out more slowly, or even in the opposite direction to the belt. In this scenario, a dynamical gap is created, across which it is harder to scatter planetesimals than in the main-sequence configuration, as discussed in \S \ref{dis-pollution}. The width of this gap can be calculated analytically, by considering the change in the width of a planet's chaotic zone to the orbit of an exterior planetesimal belt undergoing an adiabatic expansion.

The width of this chaotic zone, $\delta a_{chaos}$ is defined as: \begin{equation}
    \delta a_{\tm{chaos}}=1.3a_{\tm{pl}}\left(\frac{M_{\tm{pl}}}{M_*}\right)^{\frac{2}{7}},
\end{equation} where $M_{\tm{pl}}$ and $M_*$ are the planet and stellar mass respectively \citep{wisdom_jack_resonance_1980}. It extends from the semi-major axis of the planet ($a_{\tm{pl}}$) to $a_{\tm{pl}}\pm\delta a_{\tm{chaos}}$.  

As a star loses mass, the orbit of a planet and the outer edge of its chaotic zone and any surrounding planetesimal belts expand. The inner edge of the planetesimal belt, which is initially at the outer edge of the chaotic zone, $a_{\tm{in}}$, expands adiabatically \begin{equation}
    a_{\tm{in}}=pa_{\tm{pl}}(1+1.3\left(\frac{M_{\tm{pl}}}{M_*}\right)^{\frac{2}{7}}). \label{eq:belt}
\end{equation} The outer edge of the chaotic zone, $a_{\tm{out}}$, is affected by the altered planet-star mass ratio and the change in planet semi-major axis.
\begin{equation}
    a_{\tm{out}}=fpa_{\tm{pl}}(1+1.3\left(p\frac{M_{\tm{pl}}}{M_*}\right)^{\frac{2}{7}}).
\end{equation} If the outer edge of the chaotic zone becomes larger than the inner edge of the belt, mass from the belt ends up in the chaotic zone and becomes unstable and can be scattered inwards onto the star. However, if the planet is strongly affected by tidal forces, the outer edge of the chaotic zone expands less than the planetesimal belt and a dynamical `gap' forms in the system. The critical value of $f$ for this scenario is given in Equation \ref{eq:scattering_criterion}.
\begin{equation}
    f_{\tm{crit}}<\frac{1+1.3\left(\frac{M_{\tm{pl}}}{M_*}\right)^{\frac{2}{7}}}{1+1.3\left(p\frac{M_{\tm{pl}}}{M_*}\right)^{\frac{2}{7}}}.\label{eq:scattering_criterion}
\end{equation} In this equation $M_*$ is the initial stellar mass. For a Jupiter mass planet and $1M_\odot$ progenitor this corresponds to $f<0.975$ - an almost adiabatic migration. The points in Figure \ref{fig:summary_scatter} were classified as follows: purple, $f<0.5702$ (see Appendix \ref{app-tides}); green, $0.5702<f<f_{\tm{crit}}$; yellow, $f>f_{\tm{crit}}$ and grey, engulfed.
\begin{figure*}
    \centering
    \includegraphics[width=1\textwidth]{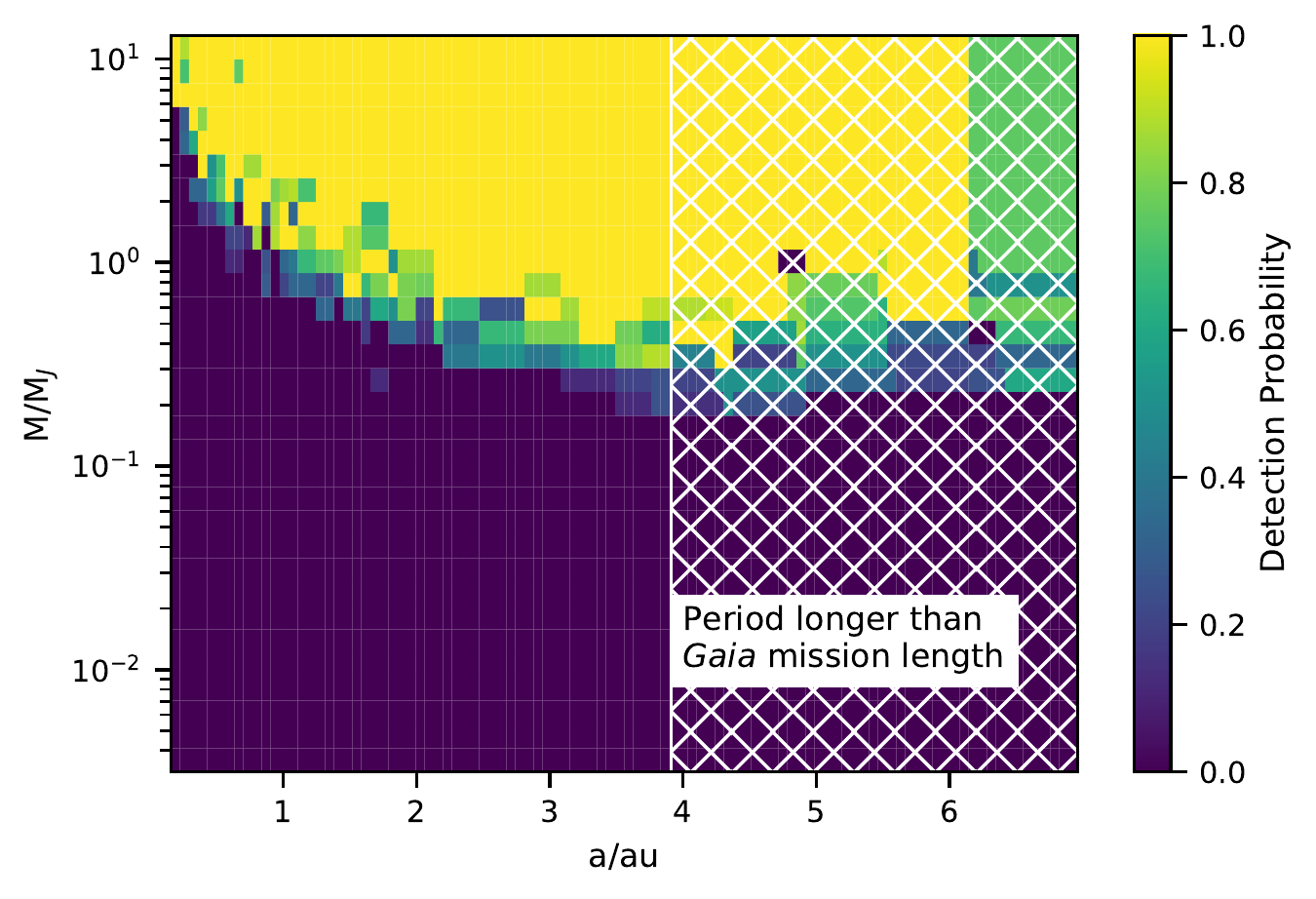}
    \caption{Detection probability, $p_{ljk}$ (\S \ref{methods-probability}) as a function of mass and semi-major axis for G29-38. Here detection probability refers to the likelihood of detecting a given planet if it exists around G29-38 and does not account for the likelihood of a given planet existing around the star.}
    \label{fig:G29_38}
\end{figure*}

\section{High detection probability white dwarfs}\label{app-table}

There are 52 white dwarfs for which the S/N of a $1\,M_J$ planet at 2\,au is greater than three (Figure \ref{fig:wd_dist}). These objects are listed in Table \ref{wd_tab}. Of particular interest is G29-38 - the first polluted white dwarf for which a debris disc was observed \citep{tokunaga_infrared_1990,jura_tidally_2003} and which has variable photospheric calcium line strengths \citep{von_hippel_discovery_2007}. \Gaia measurements of G29-38 will build on existing observations with Keck, the Hubble Space Telescope, Herschel and ALMA \citep[e.g.][]{debes_cool_2005,farihi_alma_2014,kuchner_keck_1998}. These observations currently rule out planets $>6\,M_J$ beyond 12\,au and $>16\,M_J$ between 3-12\,au \citep{debes_cool_2005}. \citet{farihi_alma_2014} use ALMA and Herschel observations to rule out the presence of dust in the 1-100\,au region emitting with $L_{IR}/L_\star>10^{-4}$ from an evolved Kuiper-belt analogue.
\Gaia could rule out the existence of planets above approximately $1\,M_J$ between 1-4\,au around this star (see Figure \ref{fig:G29_38}). A similar figure could be produced for any white dwarf in the catalogue.
This list also features WD 0141-675 around which a $\approx9\,M_J$ planet candidate has been found in DR3 \citep{arenou_gaia_2022} (see \S \ref{dis-pollution}).

\begin{table*}
\begin{tabular}[width=1\textwidth]{c|c|c|c|c|c|c|c|} \hline White dwarf name &  Gaia DR3 source ID & Distance (pc) & G$_{mag}$ & S/N for 1$M_J$ at 2\,au & error on S/N & Ca/He & Ca/H
\\\hline WDJ064509.30-164300.72 & 2947050466531873024 &       2.7 &       8.5 &                  33.7 &          5.6 &          &          \\WDJ004909.90+052318.99 & 2552928187080872832 &       4.3 &      12.3 &                  20.9 &          3.5 &  -9.92 &          \\WDJ114542.92-645029.46 & 5332606522595645952 &       4.6 &      11.4 &                  19.4 &          3.2 &          &          \\WDJ041521.80-073929.20 & 3195919254111315712 &       5.0 &       9.5 &                  18.0 &          3.0 &          &          \\WDJ043112.57+585841.29 &  470826482637310848 &       5.5 &      12.3 &                  16.3 &          2.7 &          &          \\WDJ174807.99+705235.92 & 1638979384378696704 &       6.2 &      13.8 &                  10.8 &          1.8 &          &          \\WDJ084132.43-325632.92 & 5639391810273308416 &       8.5 &      11.8 &                  10.6 &          1.8 &          & <-11.12 \\WDJ074020.79-172449.16 & 5717278911884258176 &       9.1 &      13.0 &                   9.8 &          1.6 &          &          \\WDJ031031.02-683603.38 & 4646535078125821568 &      10.4 &      11.4 &                   8.7 &          1.4 &          &          \\WDJ120526.67-233312.14 & 3489719481290397696 &      10.4 &      12.7 &                   8.6 &          1.4 &          &     -9.7 \\WDJ055509.53-041007.07 & 3022956969731332096 &       6.4 &      14.2 &                   8.6 &          1.4 &          &          \\WDJ192034.92-074000.07 & 4201781696994073472 &      10.5 &      12.3 &                   8.6 &          1.4 &          &          \\WDJ075308.14-674731.38 & 5273943488410008832 &       8.2 &      13.8 &                   8.2 &          1.4 &          &          \\WDJ214241.01+205958.12 & 1792830060723673472 &      11.0 &      13.2 &                   7.7 &          1.3 &          &          \\WDJ043747.41-084910.62 & 3186021141200137472 &       9.4 &      13.6 &                   7.6 &          1.3 &          &          \\WDJ055625.46+052148.44 & 3320184202856435840 &       8.1 &      14.0 &                   7.6 &          1.3 &          &          \\WDJ013759.39-045944.67 & 2480523216087975040 &      12.6 &      12.7 &                   7.1 &          1.2 &          & <-11.53 \\WDJ014300.98-671830.35* & 4698424845771339520* &       9.7 &      13.7 &                   7.1 &          1.2 &          &          \\WDJ133631.85+034045.94 & 3713594960831605760 &       8.3 &      14.4 &                   6.2 &          1.0 &          &          \\WDJ195629.23-010232.67 & 4235280071072332672 &      11.6 &      13.6 &                   6.2 &          1.0 &          & <-11.27 \\WDJ215140.11+591734.85 & 2202703050401536000 &       8.5 &      14.4 &                   6.2 &          1.0 &          &          \\WDJ203421.89+250349.75 & 1831553382794173824 &      14.8 &      11.5 &                   6.1 &          1.0 &          &          \\WDJ004126.03-222102.29 & 2349916559152267008 &       9.1 &      14.3 &                   5.9 &          1.0 &          &          \\WDJ154730.02-375508.46 & 6009537829925128064 &      15.2 &      13.0 &                   5.9 &          1.0 &          & <-10.28 \\WDJ133013.64-083429.47 & 3630035787972473600 &      16.1 &      12.4 &                   5.6 &          0.9 &          &          \\WDJ201056.85-301306.63 & 6749419923164242816 &      16.2 &      12.3 &                   5.6 &          0.9 &          &  <-8.84 \\WDJ015202.96+470006.66 &  356922880493142016 &      16.5 &      12.5 &                   5.5 &          0.9 &          &          \\WDJ211856.26+541241.24 & 2176116580055936512 &      17.3 &      12.4 &                   5.2 &          0.9 &          &          \\WDJ001214.75+502520.74 &  395234439752169344 &      10.9 &      14.2 &                   5.1 &          0.8 &          & <-11.15 \\WDJ113430.48-325002.40 & 3478127467639543296 &       9.6 &      14.5 &                   5.1 &          0.8 &          &          \\WDJ232847.64+051454.24 & 2660358032257156736 &      17.5 &      13.1 &                   5.0 &          0.8 &          &    -6.58 \\WDJ041630.04-591757.19 & 4678664766393827328 &      18.3 &      12.5 &                   4.9 &          0.8 &          &          \\WDJ234350.72+323246.73 & 2871730307948650368 &      18.6 &      13.0 &                   4.8 &          0.8 &          &  <-9.53 \\WDJ193713.75+274318.74 & 2025389380082340992 &      18.2 &      13.1 &                   4.8 &          0.8 &          &          \\WDJ081227.07-352943.32 & 5544743925212648320 &      11.2 &      14.3 &                   4.7 &          0.8 &          &          \\WDJ055119.48-001021.11 & 3218697767783768320 &      11.2 &      14.4 &                   4.5 &          0.8 &          &          \\WDJ211316.85-814912.88 & 6348672845649310464 &      16.2 &      13.6 &                   4.4 &          0.7 &          &     -8.6 \\WDJ233850.74-074119.97 & 2439184705619919488 &      18.6 &      13.3 &                   4.3 &          0.7 &          &          \\WDJ191858.63+384321.48 & 2052891361294411520 &      11.9 &      14.5 &                   4.2 &          0.7 &          &          \\WDJ162825.00+364615.85 & 1331106782752978688 &      15.9 &      13.8 &                   4.2 &          0.7 &  -9.07 &          \\WDJ204234.75-200435.94 & 6857939315643803776 &      21.8 &      12.4 &                   4.1 &          0.7 &          &          \\WDJ202025.46-302714.65 & 6797171060323993728 &      17.4 &      13.6 &                   4.1 &          0.7 &          &          \\WDJ122642.02-661218.47 & 5860131207828395648 &      15.4 &      13.9 &                   4.1 &          0.7 &          &          \\WDJ021228.98-080411.00 & 2486388560866377856 &      16.7 &      13.7 &                   4.1 &          0.7 &          &          \\WDJ212657.66+733844.66 & 2274076297221555968 &      22.2 &      12.9 &                   4.1 &          0.7 &          &          \\WDJ215225.38+022319.58 & 2693940725141960192 &      22.5 &      12.8 &                   4.0 &          0.7 &          &          \\WDJ011800.08+161020.56 & 2591754107321120896 &      16.8 &      13.8 &                   4.0 &          0.7 &          &          \\WDJ112412.97+212135.57 & 3978879594463300992 &      14.7 &      14.1 &                   4.0 &          0.7 &          &          \\WDJ225353.39-064654.49 & 2611561706216413696 &       8.5 &      15.4 &                   3.8 &          0.6 & -10.00 &          \\WDJ143307.64-812014.13 & 5772718006135360128 &      20.9 &      13.4 &                   3.7 &          0.6 &          &          \\WDJ080653.75-661816.70 & 5274517467840296832 &      19.2 &      13.7 &                   3.6 &          0.6 &          &          \\WDJ015151.14+642552.55 &  518201792978858880 &      17.3 &      13.9 &                   3.6 &          0.6 &          & <-12.08 
\\ \hline 
\end{tabular}
\caption{White dwarfs in the EDR3 catalogue for which a 1\,$M_J$ planet at 2\,au would have a $S/N +\sigma(s/N)>3$. The data in the first four columns come from \citet{gentilefusillo_catalogue_2021}. The white dwarf names are WD J $+$ J2000 RA (hh mm ss.ss) $+$ Dec. (dd mm ss.s), equinox and epoch 2000, the distance is the median of the geometric distance posterior (pc) \citep{bailer-jones_estimating_2021} and $G_{\rm{mag}}$ is the corrected photometric G mean magnitude \citep{brown_gaia_2021}. The S/N and error on the S/N were calculated by the authors. The Ca/H and Ca/He data comes from the Montreal White Dwarf Database \citep{dufour_montreal_2017}. The asterisk marks WD 0141-675, the white dwarf with a $\approx9\,M_J$ planet candidate in DR3 \citep{arenou_gaia_2022}.  This data along with $P_{\tm{WD}}$ and the white dwarf mass will be made available on Vizier after publication.}
\label{wd_tab}
\end{table*}



\bsp	
\label{lastpage}
\end{document}